\documentclass[prb,twocolumn,amsmath,amssymb,superscriptaddress]{revtex4}

\usepackage{graphicx}
\usepackage{xcolor}

\date{\today}                  

\begin{document}

\title{Electronic transport in one-dimensional Floquet topological insulators via topological and nontopological edge states}

\author{Niclas M\"uller}
\email[Email: ]{nmueller@physik.rwth-aachen.de}
\affiliation{Institut f\"ur Theorie der Statistischen Physik, RWTH Aachen, 
52074 Aachen, Germany and JARA - Fundamentals of Future Information Technology}
\author{Dante M. Kennes}
\affiliation{Institut f\"ur Theorie der Statistischen Physik, RWTH Aachen, 
52074 Aachen, Germany and JARA - Fundamentals of Future Information Technology}
\author{Jelena Klinovaja}
\affiliation{Department of Physics, University of Basel, Klingelbergstrasse 82, CH-4056 Basel, Switzerland}
\author{Daniel Loss}
\affiliation{Department of Physics, University of Basel, Klingelbergstrasse 82, CH-4056 Basel, Switzerland}
\author{Herbert Schoeller}
\affiliation{Institut f\"ur Theorie der Statistischen Physik, RWTH Aachen, 
52074 Aachen, Germany and JARA - Fundamentals of Future Information Technology}

\begin{abstract}
{Based on probing electronic transport properties we propose an experimental test for the recently discovered rich topological phase diagram of one-dimensional Floquet topological insulators with Rashba spin-orbit interaction [Kennes \emph{et al.}, Phys. Rev. B {\bf 100}, 041103(R) (2019)]. Using the Keldysh-Floquet formalism we compute electronic transport properties of these nanowires, where we propose to couple the leads in such a way, as to primarily address electronic states with a large weight at one edge of the system. By tuning the Fermi energy of the leads to the center of the topological gap we are able to directly address the topological edge states, granting experimental access to the topological phase diagram. Surprisingly, when tuning the lead Fermi energy to special values in the bulk which coincide with extremal points of the dispersion relation, we find additional peaks of similar magnitude to those caused by the topological edge states. These peaks reveal the presence of continua of states centered around aforementioned extremal points whose wavefunctions are linear combinations of delocalized bulk states and exponentially localized edge states, where the ratio of edge- to bulk-state amplitude is maximal at the extremal point of the dispersion. We discuss the transport properties of these \emph{non-topological edge states}, explain their emergence in terms of an intuitive yet quantitative physical picture and discuss their relationship with Van Hove singularities in the bulk of the system. The mechanism giving rise to these states is not specific to the model we consider here, suggesting that they may be present in a wide class of one-dimensional systems.}

\end{abstract}

\maketitle

\section{Introduction}
\label{sec:introduction}
One of the main goals in the field of topological phases of matter is to connect the abstract notion of topological invariants with physical observables, as e.g.~in the quantum Hall effect (QHE) [\onlinecite{klitzing_80},\onlinecite{thouless_82}] where the Hall conductance can be linked to the integer (TKNN) topological invariant. Floquet topological insulators (FTIs) are periodically driven systems in which nontrivial topological properties of the bandstructure can be induced by the drive, allowing for a high degree of controllability. Such systems have been classified [\onlinecite{kitagawa_etal_10}-\onlinecite{hoeckendorf_etal_prb_18}] and experimentally realized in photonic crystals [\onlinecite{kitagawa_etal_natcomm_12}-\onlinecite{mukherjee_etal_natcomm_17}], cold atom systems [\onlinecite{jotzu_etal_nature_14}-\onlinecite{quelle_etal_njp_17}], and solid-state materials [\onlinecite{calvo_etal_apl_11}-\onlinecite{wang_scirep_17}]. Among the predicted topological properties are the emergence of Majorana edge modes [\onlinecite{jiang_etal_prl_11,kundu_seradjeh_prl_13,thakurathi_etal_prb_13,thakurathi_sengupta_sen_prb_14}] and parafermions [\onlinecite{thakurathi_loss_klinovaja_prb_17}] in one-dimensional (1D) FTIs, the photoinduced QHE in 2D materials [\onlinecite{oka_aoki_prb_09}-\onlinecite{perez-piskunow_etal_pra_15}], topological surface states in 3D FTIs [\onlinecite{lindner_etal_prb_13}] and Weyl semimetals and fractional FTIs in coupled Rashba nanowires [\onlinecite{klinovaja_stano_loss_prl_16}]. 

{In the context of FTIs the topological invariant, which is associated with the topology of the quasienergyspectrum of the Floquet Hamiltonian, is not as \emph{meaningful} as e.g.~in the QHE case, as it is sensitive to the truncation of the Floquet Hamiltonian in Floquet space. If the driving frequency $\omega$ is small compared to the bandwidth $W$ of the static system (as is often the case in a solid-state context) a large number (of order $\sim\frac{W}{\omega}$) of overlapping, coupled Floquet replicas has to be taken into account in describing the physics near the topological gap at zero energy. This multitude of quasienergy bands potentially leads to a rich topological phase diagram with a large number of degenerate topological edge states (TESs) [\onlinecite{kennes_19}]. As long as the quasienergy bands of an additional $l-$th Floquet replica extend into the initial gap (i.e.~if $|l|\omega\lesssim W$) the additional quasienergy bands lead to additional anti-crossings in the quasienergyspectrum at increasing quasimomenta, which can potentially host TESs. Thus, by increasing the truncation order, the degenerate subspace of TESs, or equivalently the value of the topological invariant, grows. Since the gaps associated with anti-crossings at increasing quasimomenta decrease rapidly in size, the corresponding localization lengths (which are inversely proportional to the gap size) increase, leading to a hierarchy of TESs of drastically different \emph{qualitiy} with regard to observables which measure the local density of states (LDOS) at the edge of the system. The mere number of TESs, i.e.~the topological invariant however utterly misses this point and thus tells an incomplete story. }

In order to specifically address edge states with localization lengths on a scale $\lesssim L$, we propose to couple the leads only to the edge of the wire on a scale $L$. In doing so, we effectively probe the LDOS at the edge of the nanowire, which is large if well-localized edge states are present. This gives experimental access to the predicted topological phase diagram, which does not only contain information on the number of TESs at any given truncation order in Floquet space, but is also supplemented by the corresponding localization lengths [\onlinecite{kennes_19}]. 

Edge state transport in FTIs has been studied using effective Floquet Boltzmann methods (taking into account occupations of TESs) in 2D systems [\onlinecite{esin_18}], Floquet-Green's function methods in 2D [\onlinecite{kitagawa_11}-\onlinecite{huaman_19}] and 1D systems [\onlinecite{kundu_seradjeh_prl_13}] and Floquet scattering theory [\onlinecite{torres_14}] in 2D systems. In contrast to these works, we compute the full ac-conductance (not just the dc-component) and we consider system sizes which are realistic for solid-state implementations. {It is only on these scales that the finite-size effects associated with \emph{non-topological edge states} (which are discussed further below) become negligible and that significant contributions from these states can be observed in transport simulations. }

{The intrinsic non-equilibrium nature of FTIs poses many challenges. One such challenge is the fact that the occupation of Floquet states shows a high degree of dependence on details of fermionic and bosonic baths coupled to the system [\onlinecite{esin_18},\onlinecite{seetharam_15}-\onlinecite{deghani_14}]. We circumvent these difficulties by concentrating on observables that do not depend on the occupation of states, but rather just on the spectral properties or LDOS [\onlinecite{kundu_seradjeh_prl_13},\onlinecite{kitagawa_11}-\onlinecite{huaman_19}], which allows for a straightforward inquiry into the topological properties. Assuming the wide-band limit for the reservoirs we find that the differential conductance (even for the driven case) solely depends on the LDOS and is thus a convenient observable for our purposes. }

{A topic that has gained little attention in the context of topological phases of matter (or, more generally, the physics of boundaries of crystalline materials) concerns \emph{non-topological edge states} (NTESs). These novel states live at energies which overlap with the bulk bands and are linear combinations of exponentially localized edge states and delocalized bulk states, where the amplitude of the edge state component is peaked near extremal points of the quasienergy dispersion, leading to an excess density at the edge relative to the bulk-average. Here, we report the discovery of such states in the model under consideration. In the bulk the extremal points at which NTESs arise are associated with van Hove singularities (VHSs), leading to a large bulk-LDOS. At the edges however, where the system is no longer translationally invariant, the peaks in the LDOS due to VHSs smoothly pass over into peaks caused by the edge state component of the NTESs. Clearly separating the effect of NTESs (which are a pure edge effect) from the effect of the VHSs (which are a pure bulk effect) is only possible, if the length scale on which the edge state component decays is smaller than the length scale on which the VHS manifests itself, i.e.~the length scale on which the system \emph{realizes} that it is translationally invariant.}

{Predicting the existence and analyzing the properties of NTESs numerically is challenging, because unlike TESs they are not linked to topological invariants. Furthermore, since these states live near extremal points of the dispersion where the energy crosses the bands at arbitrarily small quasimomenta $k$, the bulk contribution of these states oscillates at arbitrarily long length scales $\frac{1}{k}$, making NTESs especially prone to finite-size effects if the system size $N$ does not exceed these large length scales. We overcome this challenge by employing an efficient algorithm with complexity $\mathcal{O}(N)$ (instead of the usual $\mathcal{O}(N^{2\dots3})$) for the computation of the differential conductance in the framework of the Keldysh-Floquet formalism, which allows us to treat extremely large system sizes of up to $N=10^7$ unit cells.}

{In order to reach a better understanding of the NTESs we analytically construct and analyze these states for the case of the lowest truncation of the Floquet Hamiltonian in the half-infinite limit, employing the same method that was used in determining the topological phase diagram of the system under consideration [\onlinecite{kennes_19}]. We argue how this construction can be generalized to higher truncations, explaining the existence of NTESs in terms of an intuitive yet quantitative picture involving extremal points of the dispersion relation in the complex quasimomentum plane. We discuss the contribution of NTESs to the transport properties of the nanowire and thereby propose a method for observing them experimentally. Finally, we state those properties of the model under consideration which are responsible for the emergence of NTESs, opening a route for the systematic investigation of other models with regard to NTESs. }

\section{Model and experimental setup}
\label{sec:model_and_experimental_setup}

In this section we present the models of both the driven Rashba nanowire and the leads that carry current to and from the wire. Using the wide-band limit and Keldysh-Green's functions, we compute the self-energy contribution of the leads to the wire Hamiltonian. The sum of the Hamiltonian and the self-energy defines the \emph{effective} Hamiltonian, from which the spectral/transport properties are computed. If the scale of the lead-induced self-energy is small compared to the topological gap, the topological edge states are unaltered by this coupling and thus the topological properties of the wire Hamiltonian also manifest in the effective Hamiltonian.

\subsection{Wire Hamiltonian}
\label{sec:wire}

The model we consider here was introduced in [\onlinecite{thakurathi_loss_klinovaja_prb_17}] and a detailed study of its topological properties was conducted in [\onlinecite{kennes_19}]. This one-dimensional two-band Rashba nanowire features TESs if a transverse Zeeman field $\Delta_Z$ and a periodic drive of the inter-band transition $t_F$ is applied. Denoting Pauli-matrices in band/spin space by $\eta_i/\sigma_i$ respectively the single-particle Hamiltonian in quasimomentum space reads
\begin{equation}
\label{eq:1}
h_k(t) = (\epsilon_k + \alpha \sin k \, \sigma_z)\eta_z + \Delta_Z \sigma_x + 2t_F \cos \omega t \, \eta_x,
\end{equation}
where $\epsilon_k = E_k + \frac{\Delta_g}{2}$ and $E_k=W\sin^2 \frac{k}{2}$. Here, $W$ denotes the bandwidth, $-\pi<k\leq \pi$ is the quasimomentum, $\Delta_g$ is the (intrinsic) bandgap (not to be confused with the topological gap $\Delta$ that emerges in the quasienergyspectrum of the Floquet Hamiltonian), $\alpha$ is the Rashba constant and $\omega$ denotes the driving frequency. Such a model could be realized in curved bilayer graphene, where both the bandgap and strength of the Rashba coupling can be controlled [\onlinecite{klinovaja_12}]. 

We perform the following exact unitary transformation in order to work in the rotating wave approximation (RWA) basis
\begin{equation}
h_k(t) \rightarrow \bar{h}_k(t) = U^{\dagger}(t) h_k(t) U(t) - i U^{\dagger}(t) \dot{U}(t) ,
\end{equation} 
where $U(t) = e^{-i\frac{\omega t}{2}\eta_z} = \sum_{\eta=\pm} P_{\eta} e^{-i\eta\frac{\omega t}{2}}$ and $P_{\eta} = \frac{1+\eta \eta_z}{2}$ is a projector on band $\eta$. At resonance ($\omega=\Delta_g$) the transformed Hamiltonian reads
\begin{equation}
\label{eq:2}
\bar{h}_k(t) = h^R_k + t_F (e^{i\Omega t} \eta_+ + e^{-i\Omega t} \eta_-),
\end{equation}
where we have defined the RWA Hamiltonian
\begin{equation}
h^R_k = (E_k + \alpha \sin k \, \sigma_z)\, \eta_z + \Delta_Z \sigma_x + t_F \eta_x,
\end{equation}
the effective driving frequency $\Omega=2\omega$ and the raising/lowering operators $\eta_{\pm} = \frac{\eta_x \pm i \eta_y}{2}$. The corresponding Floquet Hamiltonian reads
\begin{equation}
\label{eq:3}
(\bar{h}^F_k)_{ll^{\prime}} = (h^R_k - l\omega)\delta_{ll^{\prime}} + t_F (\delta_{l,l^{\prime}-2} \, \eta_+ + \delta_{l,l^{\prime}+2} \, \eta_-),
\end{equation}
where $l,l^{\prime} \in \mathbb{Z}$. In the following we truncate this Hamiltonian symmetrically in Floquet space at $|l|,|l^{\prime}|\leq l_m$ and increase $l_m$ until we find convergence in the observables that are computed from the effective Floquet Hamiltonian. Furthermore, we transform back from momentum to real space by substituting $e^{\pm ik}\rightarrow \delta_{n,n^{\prime}\pm1}$ to obtain a $[4(2l_m+1)N]$-dimensional tight-binding Hamiltonian with $N$ unit cells and open boundary conditions. Note that we set the lattice spacing to unity here. 

Finally, we note some properties of the Floquet Hamiltonian defined in Eq.~(\ref{eq:3}): it features a local chiral symmetry $S$ for every (symmetric) truncation $-l_m\leq l,l^{\prime} \leq l_m$, i.e.~$S \bar{h}^F_k S = - \bar{h}^F_k$ with the chiral symmetry operator $S_{ll^{\prime}} = \delta_{\bar{l}l^{\prime}} \eta_y \sigma_z$, where $\bar{l}=-l$, and is thus a representative of the BDI symmetry class [\onlinecite{kitagawa_etal_10}-\onlinecite{hoeckendorf_etal_prb_18}]. Also note, that this Hamiltonian can be decomposed into a sum of two commuting parts, acting only on the odd and even Floquet indices respectively. Both of these properties are crucial in determining the topological properties and computing the topological invariant of the system. Here, however, the coupling to leads destroys both of these features. Finally, we note that the Rashba-term breaks inversion-symmetry, i.e.~$\bar{h}^F_{-k} = \sigma_x \bar{h}^F_k \sigma_x \neq \bar{h}^F_k$. {As we will see in Appendix~\ref{sec:ntes_rwa}, this property is crucial for the emergence of non-topological edge states.}

\subsection{Reservoirs and self-energy}
\label{sec:reservoirs}

\begin{figure}
\includegraphics[scale=0.55]{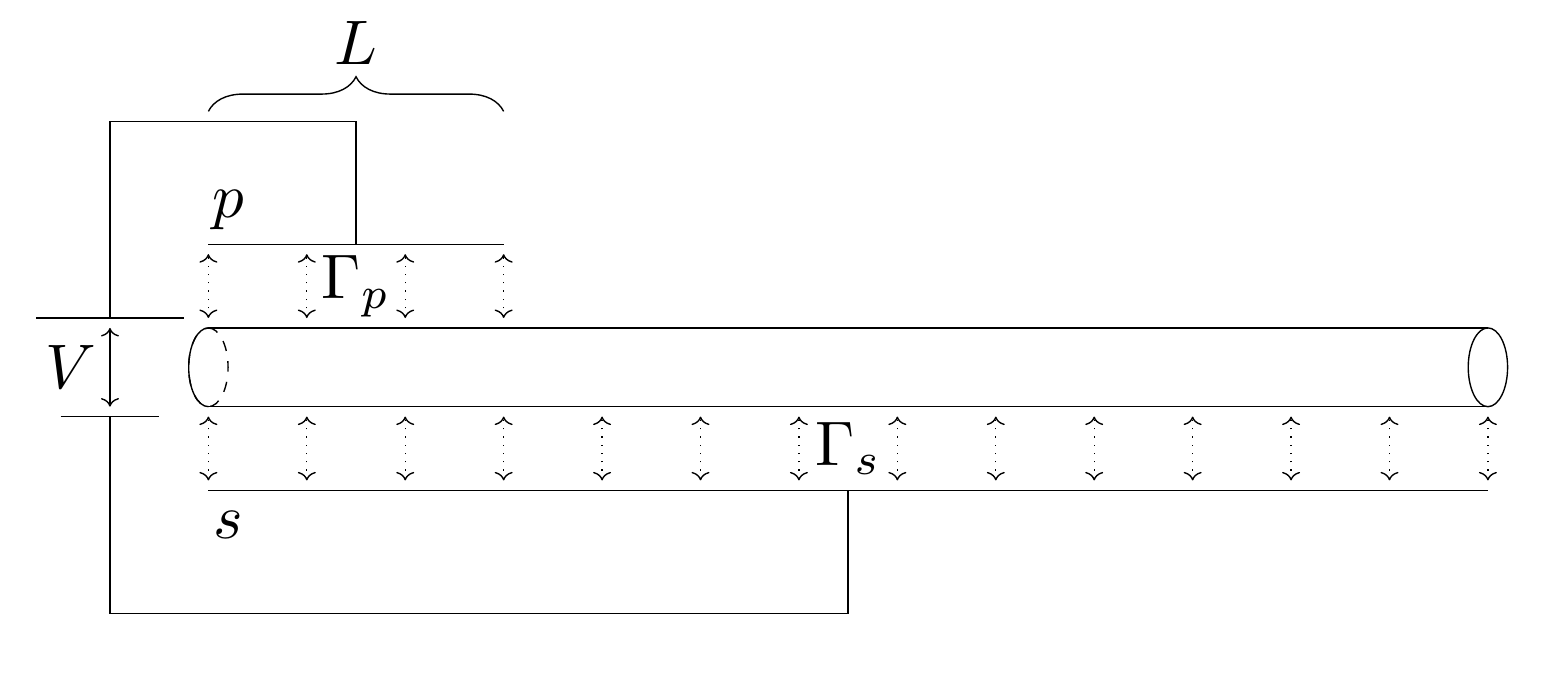}
\caption{Schematic diagram depicting the experimental setup. A bias voltage $V$ is applied between the two reservoirs, the probe (p) which couples only to the edge of the wire on a scale $L$ and the substrate (s), which couples to the whole wire. The microscopic details of the coupling between the reservoirs and the wire are encoded in the hybridization functions $\Gamma_{p/s}$ respectively. \label{fig:setup_diagram}}
\end{figure}

We aim at probing the spectral properties of the Hamiltonian introduced in the previous section via electronic transport measurements. It is thus necessary to couple reservoirs to the wire which serve as leads for the current flowing to and from the system. The setup we have in mind is depicted in Fig.~\ref{fig:setup_diagram}: we consider two reservoirs, one is called the probe (p) and the other the substrate (s) in the following. In order to induce a current these reservoirs are kept at chemical potentials $\mu_s = \mu_0$ and $\mu_p=\mu_0+eV$ with electric charge $e$, voltage difference $V$ and lead Fermi energy $\mu_0$. {Additionally, we assume the zero temperature and wide-band limits for both reservoirs.} Note that this setup is unconventional in the sense that the transport does not occur along the wire, but rather via its edge. 

{All (renormalized) single-particle states of the wire serve as channels for the current. We are however exclusively interested in the signal due to topological (and non-topological) edge states. These states are characterized by the fact that a significant portion of the weight of their wavefunctions is concentrated near the edges of the wire. If the probe only couples to the first $L\ll N$ unit cells, the contribution of delocalized states to the differential conductance is suppressed by a factor ${L}/{N}$ compared to exponentially localized states (at least as long as the localization length is smaller than $L$). Thus, in contrast to other works on transport in FTIs, here, delocalized states will only contribute an insignificant background to the total differential conductance. }

{In our simulations we assume that the coupling between probe and wire is much weaker than the substrate-wire coupling, which itself is taken to be much smaller than the smallest energy scale of the model. This so-called weak-coupling approximation (WCA) is numerically advantageous, because, in combination with a suitable choice for the spatial structure of the hybridization functions $\Gamma_{p/s}$ discussed below, it leaves the differential conductance independent of off-diagonal elements (in position space) of the Green's functions, which in turn allows for the application of aforementioned efficient algorithm in the computation of the conductance. Relaxing this approximation excludes the possibility to treat those large system sizes which are necessary for the observation of NTESs. The WCA is used frequently in the context of transport simulations, where back action of the probe lead (e.g.~scanning tunneling/probe microscopes) on the device are of no interest and should therefore be avoided (in contrast to the substrate reservoirs, which are responsible for tuning the occupations in the device). Nonetheless, in Appendix~\ref{sec:beyond_WCA} we give an estimate of the corrections beyond WCA and argue that they do not qualitatively change the results, even if one relaxes the WCA.}

In order to describe the effect of the coupling between the wire and the reservoirs, we refer to the Keldysh-Floquet formalism (see Appendix~\ref{sec:keldysh-floquet_formalism} for a brief review), in which the influence of the reservoirs is encoded in the reservoir-induced self-energy $\Sigma = \Sigma_s + \Sigma_p$, with individual contributions from the substrate and probe lead respectively. In the following we derive expressions for a single reservoir, dropping the $s/p$ index for notational convenience. 
 
 Assuming the reservoirs to be non-interacting and static, i.e.~not affected by the external driving, the diagrammatic expansion of the self-energy is given by $\Sigma(t,t^{\prime}) = \mathcal{V}^{\dagger} g(t,t^{\prime}) \mathcal{V}$, where the matrix elements of the vertex $\mathcal{V}$ describe the overlap between single-particle states of reservoir and wire respectively, and where $g$ denotes the (non-interacting) Green's function of the reservoir. We define the hybridization function as $\Gamma(E) = 2\pi \mathcal{V}^{\dagger} \delta(E-h)\mathcal{V}$ where $h$ denotes the single-particle Hamiltonian of the reservoir. Since we are interested in the universal physics irrespective of microscopic details of the reservoirs and the reservoir-wire couplings, we employ the wide-band limit (WBL), i.e.~we assume $\Gamma$ to be independent of the energy argument, which is reasonable as long as the spectrum of the reservoir varies slowly on the scale on which $E$ is evaluated. 
 
Using the definitions of the non-interacting Green's functions defined in Eqs.~(\ref{eq:A1.13})-(\ref{eq:A1.15}) and assuming the WBL we find the following expressions for the retarded, lesser and greater self-energy components:
\begin{align}
\label{eq:4a}
\Sigma^R(t,t^{\prime}) &= -\frac{i}{2} \Gamma \delta(t-t^{\prime}), \\
\label{eq:5}
\Sigma^<(t,t^{\prime}) &= i \Gamma \int \frac{\mathrm{d}E}{2\pi} e^{-iE(t-t^{\prime}+i0^+)} \Theta(\mu-E), \\
\label{eq:6}
\Sigma^>(t,t^{\prime}) &= i \Gamma \int \frac{\mathrm{d}E}{2\pi} e^{-iE(t-t^{\prime}-i0^+)} \left[\Theta(\mu-E)-1\right],
\end{align}
where we have replaced the Fermi-functions with step-functions $\Theta$ due to the zero-temperature limit and where $e^{\pm E0^+}$ are convergence factors. 

We anticipate that topological properties of the effective Hamiltonian will be lost, if the scale of $\Gamma$ becomes comparable to the scale of the topological gap $\Delta$ and thus assume $\Gamma \ll \Delta$ in the following. This can of course only be ensured locally in parameter space, since the gap becomes arbitrarily small near the phase boundaries. {In order to still achieve a smooth density of states we choose the scale of $\Gamma$ to be of the order of the level spacing $\delta_{\epsilon}$. Note that the bandwidth of the quasienergyspectrum is given by $\approx 2W + 2l_{m}\omega$, while the number of states is given by $4N(2l_m +1)$, such that the level spacing (for a sufficiently large truncation) is given by $\delta_{\epsilon} \sim \frac{\omega}{N}$, not $\sim \frac{W}{N}$.} For the large system sizes we are considering this scale is tiny, such that the detailed structure of $\Gamma$ in band-, spin- and position space does not have a strong influence on the results presented in the following. We thus choose a particularly convenient structure, where the hybridization function is diagonal in position space, i.e.~does not induce hopping between distinct unit cells $\left<n\left|\Gamma\right|n^{\prime}\right> \sim \delta_{nn^{\prime}}$. {As we will see, this property of the hybridization function is crucial with regard to the efficient ($\mathcal{O}(N)$) numerical evaluation of the differential conductance (see Sec.~\ref{subsec:inversion_algorithm})}. For the structure in band-spin space we consider two alternatives, the first being a uniform coupling of both bands and spin components $\left<\eta \sigma\left|\Gamma\right|\eta^{\prime}\sigma^{\prime}\right> \sim 1$ and the second being a diagonal structure in these subspaces $\left<\eta \sigma\left|\Gamma\right|\eta^{\prime}\sigma^{\prime}\right> \sim \delta_{\eta \eta^{\prime}}\delta_{\sigma\sigma^{\prime}}$. Any experimental configuration should lie somewhere between these two limiting cases.

\section{Methods}
\label{sec:methods}

{In this section we present the methods employed in the computation of the observables, which in turn are used in order to probe the LDOS at the edge of the nanowire.} We give a derivation of the expression for the current flowing through the edge of the wire in terms of Floquet Green's functions, assuming the geometry introduced in the previous section. Furthermore, we present the algorithm used for the computation of the Green's functions, where we exploit the aforementioned diagonal structure of the hybridization function in position space {and the weak-coupling approximation} in order to greatly reduce the computational effort and allow for the efficient treatment of extremely large system sizes of up to $10^7$ unit cells.

\subsection{Differential conductance formula}
\label{sec:current_formula}

The current flowing from the probe reservoir into the wire can be expressed as
\begin{align}
\nonumber
&I(t) = -e \int \displaylimits_{-\infty}^t \mathrm{d}t^{\prime} \, \mathrm{Tr} \\
&\times \{\Sigma_p^>(t,t^{\prime}) G^<(t^{\prime},t)-\Sigma_p^<(t,t^{\prime}) G^>(t^{\prime},t) + (t\leftrightarrow t^{\prime})\},
\end{align}
where $\mathrm{Tr}$ denotes the trace over all single-particle states in the wire, $\Sigma^{\gtrless}_p$ is the probe-induced greater/lesser self-energy and $G^{\gtrless}$ is the fully dressed greater/lesser wire Green's function. Employing the weak coupling approximation (WCA) we assume that the substrate-induced self-energy is much larger than the probe-induced self-energy, i.e.~we neglect the contribution of $\Sigma_p$ in $G^{\gtrless}$. {At this level of approximation the Green's functions are independent of the bias voltage $V$ and thus the occupation of single-particle (Floquet) states in the wire is solely controlled by the substrate. In Appendix~\ref{sec:beyond_WCA} we compute corrections beyond the WCA and argue that they do not qualitatively change the results, while introducing a dependence of the differential conductance on off-diagonal terms (in position space) of the Green's functions, even if the hybridization functions do not have such terms.}

From Eqs.~(\ref{eq:5}) and (\ref{eq:6}) we find
\begin{equation}
\label{eq:5a}
\partial_V \Sigma_p^{\gtrless}(t,t^{\prime})\rvert_{V\rightarrow 0} = \frac{ie\Gamma_p}{2\pi} e^{-i\mu_0(t-t^{\prime})}.
\end{equation}
Denoting the differential conductance at zero bias voltage in units of the conductance quantum $\frac{e^2}{\pi}=\frac{2e^2}{h}$ as $g(t)$ we find
\begin{align}
\nonumber
&g(t) \equiv \frac{\pi}{e^2}\partial_V I(t)\rvert_{V\rightarrow 0} = - \int\displaylimits_{-\infty}^t\mathrm{d}t^{\prime} \, \mathrm{Tr} \, \Gamma_p \\
&\times \{e^{i\mu_0(t-t^{\prime})} \frac{G^>(t,t^{\prime})-G^<(t,t^{\prime})}{2i} + (t\leftrightarrow t^{\prime})\} \\
\label{eq:4}
&=-\Im \int\displaylimits_{-\infty}^t\mathrm{d}t^{\prime} \,e^{i\mu_0(t-t^{\prime})}\, \mathrm{Tr} \, \Gamma_p G^R(t,t^{\prime}),
\end{align}
where we have used the identities in Eqs.~(\ref{eq:A1.9}) and (\ref{eq:A1.10}), where $G^R$ denotes the retarded Green's function, which only depends on the spectral properties of the substrate-dressed wire and where we have used natural units, setting Planck's constant $\hbar = \frac{h}{2\pi}$ to unity. 

Next, we transform the retarded Green's function into the RWA basis ($G^R(t,t^{\prime}) = U(t) \bar{G}^R(t,t^{\prime}) U^{\dagger}(t^{\prime})$) and substitute this form into Eq.~(\ref{eq:4}). Using the cyclic property of the trace we find
\begin{align}
\nonumber
g(t) = -\Im \mathrm{Tr} \sum_{\eta=\pm} \left(P_{\eta} \Gamma_p P_{\eta} + e^{i\eta \omega t} P_{\eta} \Gamma_p P_{\bar{\eta}}\right) \\
\times \int\displaylimits_{-\infty}^t \mathrm{d}t^{\prime} e^{i\mu_{\eta}(t-t^{\prime})} \bar{G}^R(t,t^{\prime}),
\end{align}
where $\bar{\eta} = -\eta$ and where we have defined the effective chemical potential $\mu_{\eta} = \mu_0 -\eta \frac{\omega}{2}$. Expressing the real-time Green's function via its frequency-Floquet space analogue (see Eq.~(\ref{eq:A1.22})) and performing the integration over $t^{\prime}$ we find
\begin{align}
\nonumber
&g(t) = -\Im  \sum_{l\in\mathbb{Z}} e^{-il\omega t} \sum_{\eta=\pm} \\
 &\times \mathrm{Tr} \left(P_{\eta} \Gamma_p P_{\eta} + e^{i\eta \omega t} P_{\eta} \Gamma_p P_{\bar{\eta}}\right) \bar{G}^R_{l0}(\mu_{\eta}) = \sum_{l\in\mathbb{Z}} e^{-il\omega t} g_l,
\end{align} 
where $\bar{l}=-l$ and where the components $g_l=g_{\bar{l}}^{\ast}$ are given by 
\begin{align}
\nonumber
g_l = \frac{1}{2i} \sum_{\eta=\pm} \mathrm{Tr}\Big{\{}P_{\eta} \Gamma_p P_{\eta} \left[\left(\bar{G}^R_{-l0}(\mu_{\eta})\right)^{\ast}-\bar{G}^R_{l0}(\mu_{\eta})\right] \\
\label{eq:g_l}
+ P_{\eta} \Gamma_p P_{\bar{\eta}} \left[\left(\bar{G}^R_{\eta-l,0}(\mu_{\eta})\right)^{\ast}-\bar{G}^R_{\eta+l,0}(\mu_{\eta})\right]\Big{\}}.
\end{align}
Note that, since we have assumed the probe hybridization function to be diagonal in position space, the trace only picks out the diagonal elements (in position space) of the Green's functions. It thus suffices to compute these diagonal elements, which is computationally much more feasible for the large system sizes we are aiming at. Furthermore, we would like to stress that, since the differential conductance as a function of the lead Fermi energy $\mu_0$ is given by the sum of two terms, which are symmetric with respect to $\mu_0=\pm\frac{\omega}{2}$ respectively, the chiral symmetry that is present in the Floquet Hamiltonian in the RWA basis is no longer present in this observable, i.e.~the differential conductance is neither symmetric with respect to $\mu_0-\frac{\omega}{2}\rightarrow-(\mu_0-\frac{\omega}{2})$, nor with respect to $\mu_0+\frac{\omega}{2}\rightarrow-(\mu_0+\frac{\omega}{2})$.

Finally, we give rough estimates for the bulk and TES-contribution to the differential conductance at resonance, i.e.~with the Fermi energy tuned to the gap center: Every single-particle Floquet state $\psi$ with quasienergy $\epsilon$ roughly contributes $\sim \left< \psi \left| \Gamma_p \right| \psi\right> \frac{\Gamma_s}{\left(\mu-\epsilon\right)^2 + \Gamma_s^2}$ to the total (time-averaged) differential conductance. For a TES with localization length $\ell$ we find $\left< \psi_{TES} \left| \Gamma_p \right| \psi_{TES}\right> \sim \frac{L \Gamma_p}{\ell}$, i.e.~$g_{TES}\sim \frac{L}{\ell} \frac{\Gamma_p}{\Gamma_s}$. For a (delocalized) bulk state we find $\left< \psi_{bulk} \left| \Gamma_p \right| \psi_{bulk}\right> \sim \frac{L \Gamma_p}{N}$ and $\frac{\Gamma_s}{\left(\mu-\epsilon\right)^2 + \Gamma_s^2}\sim \frac{\Gamma_s}{\epsilon^2}$. Integrating all of these contributions from the topological gap $\Delta$ up to the bandwidth $B\sim \omega l_m \gg \Delta$ (assuming a finite driving frequency $\omega$) and assuming a uniform density of states $\sim \frac{Nl_m}{B} \sim \frac{N}{\omega}$ we find $g_{bulk} \sim L \frac{\Gamma_p \Gamma_s}{\omega \Delta}$. We thus conclude, that the TES contribution to the total differential conductance at resonance dominates over the bulk contribution if $\Delta \gg \ell \frac{\Gamma_s^2}{\omega}$, which can always be achieved for small enough broadening $\Gamma_s$. Note that in this estimate $\Gamma_{p/s}$ denotes the scale of the matrix elements of the hybridization functions, not the hybridization functions themselves and that we have assumed convergence in Floquet space, i.e.~$l_m \gg \frac{W}{\omega}$.

\subsection{Inversion algorithm}
\label{subsec:inversion_algorithm}

As was shown in the previous section, in order to compute the differential conductance, we need to compute the diagonal blocks (in position space) $G^R_{nn}$ of the retarded Green's function. The retarded Floquet Green's function is given by the inverse of the matrix $T = E - \bar{h}^F + \frac{i}{2} \bar{\Gamma}_s$, where $\bar{\Gamma}_s$ denotes the substrate hybridization function in the RWA basis. Because we assumed the hybridization function to be diagonal in position space, $T$ has a block-tridiagonal structure, i.e.~$T_{nn} = a$, $T_{n,n+1} = b$, $T_{n,n-1}=b^{\dagger}$, $T_{nm}=0$ if $|n-m|>1$ and $n,m=1,\dots,N$. Here, $a,b,b^{\dagger}$ are $\left[4(2l_m+1)\right]$-dimensional matrices which can easily be read off from the definitions of the truncated Floquet Hamiltonian and the hybridization function. 

{There exist algorithms relying on an iterative scheme with an $\mathcal{O}(N)$ complexity (see e.g.~[\onlinecite{taylor_69},\onlinecite{andergassen_04}]) for computing the main diagonal of the Green's function for a 1D tridiagonal (effective) Hamiltonian with a single channel. Multichannel generalizations of this algorithm have been presented in [\onlinecite{spratlin_79},\onlinecite{sancho_85}]. In the following we will review this multichannel generalization.}

In a first step we perform a (block-)\emph{UDL} decomposition of the inverse of the Green's function, i.e.~$T=UDL$, where the matrices $U$ and $L$ take the following forms:
\begin{align}
U &= \begin{pmatrix}
1 & U_1 &                 \\
  & 1   & U_2             \\
  &     & \ddots & \ddots \\
  &     &        & \ddots & U_{N-1}\\
  &     &        &        & 1
\end{pmatrix},
\\
L &= \begin{pmatrix}
1    &     & \\
L_1  & 1   &  \\
     & L_2 & \ddots \\
     &     & \ddots & \ddots \\
     &     &        & L_{N-1} & 1
\end{pmatrix},
\end{align}
with all other matrix elements being zero. Here, $D$ is a block-diagonal matrix $D=\mathrm{diag}(\{D_i\}_{i=1\dots N})$. The ($[4(2l_m+1)]$-dimensional) matrices $U_i,L_i$ and $D_i$ can be computed in $\mathcal{O}(N)$-time given $a,b$ and $b^{\dagger}$ using the following recursion relations:
\begin{align}
\label{eq:it1}
D_N &= a, \\
\label{eq:it2}
D_{i-1} &= a - b D_i^{-1} b^{\dagger}, \\
\label{eq:U_i}
U_i &= b D_{i+1}^{-1}, \\
\label{eq:L_i}
L_i &= D_{i+1}^{-1} b^{\dagger} .
\end{align}
Next, we compute the inverses of $U$ and $L$, which are given by
\begin{equation}
U^{-1} = 
\begin{pmatrix}
1 & -U_1 & U_1U_2 & \hdots & (-1)^{N-1}U_1\dots U_{N-1} \\
  & 1    & -U_2   & \ddots & \vdots                     \\
  &      & \ddots & \ddots & U_{N-2}U_{N-1}             \\
  &      &        & \ddots & -U_{N-1}                   \\
  &      &        &        & 1
\end{pmatrix}
\end{equation}
and 
\begin{align}
\nonumber
&L^{-1} = \\
&\begin{pmatrix}
1                          &        &                &          &  \\
-L_1                       & \ddots &                &          &  \\
L_2 L_1                    & \ddots & \ddots         &          &  \\
\vdots                     & \ddots & -L_{N-2}       & 1        &  \\
(-1)^{N-1}L_{N-1}\dots L_1 & \hdots & L_{N-1}L_{N-2} & -L_{N-1} & 1 
\end{pmatrix},
\end{align}
where, again, all other matrix elements are zero. Multiplying the inverses back together we find that the diagonal blocks of $G=T^{-1}=L^{-1}D^{-1}U^{-1}$ obey the following recursion relation
\begin{align}
\label{eq:G1}
G_{11} &= D_1^{-1}, \\
\nonumber
G_{n+1,n+1} &= D_{n+1}^{-1} + L_n G_{nn} U_n \\
\label{eq:G2}
&= D_{n+1}^{-1} + D_{n+1}^{-1} b^{\dagger} G_{nn} b D_{n+1}^{-1}.
\end{align}
The algorithm thus consists of the following steps:
\begin{enumerate}
\item {Iterate Eq.~(\ref{eq:it2}) $N$ times, with initial condition determined by Eq.~(\ref{eq:it1}) and save only the last $L$ matrices $\left\{D_1,\dots, D_L\right\}$.}
\item {Using the matrices $\left\{D_1,\dots, D_L\right\}$ compute the first $L$ diagonal blocks of the Green's functions via Eqs.~(\ref{eq:G1})-(\ref{eq:G2}).}
\end{enumerate}
The complexity of this algorithm is $\mathcal{O}(N l_m^3)$ (due to step 1.), while the memory cost only scales as $\mathcal{O}(L l_m^2)$. This allows us to efficiently and numerically accurately treat extremely large system sizes of up to $N=10^7$ unit cells, which is necessary in order to rule out finite-size artifacts and to observe significant signals from the non-topological edge states.

\section{Results and Discussion}

In this section, we analyze the differential conductance defined in the previous section. Our goal here is twofold: (a) we want to extract the topological phase diagram, i.e.~address the TESs which live in the topological gap and (b) probe the system at finite energies relative to the gap center, which reveals the NTESs. 

In order to achieve goal (a) we tune the Fermi energy to the center of the topological gap, vary the driving amplitude and Zeeman energy, and compute the differential conductance in this two-dimensional phase space. We compare these results to the previously predicted topological phase diagram (see [\onlinecite{kennes_19}]). In a second step, in order to achieve goal (b), we keep the driving amplitude constant and vary the Fermi level and Zeeman energy, probing NTESs at finite energies relative to the gap center. We discuss the properties of the NTESs and explain their emergence in terms of an intuitive yet quantitative picture. 

All simulations are performed for $N=10^7$ unit cells and with a cutoff in Floquet space at $l_m=5$, at which point we find convergence in all quantities to the percent regime. We set $\Gamma_s = 10^{-3} \, \Omega$ and $L=500$. For the bandwidth and Rashba constant we choose $W = 20 \, \Omega$ and $\alpha=\frac{3}{4}\Omega$.

To enhance the fine structure that is present in the data we present results in terms of the quantity
\begin{equation}
\mathcal{L} g_l \equiv \log_{10}\left(\frac{\Gamma_s}{\Gamma_p}|g_l|\right),
\end{equation} 
i.e.~we first rescale the absolute value of the Fourier components of the differential conductance by the ratio of substrate to probe broadening, such that the contribution of a well localized state at resonance is of order unity. Subsequently, we take the logarithm of this quantity. This is convenient since the (scaled) differential conductance varies over several orders of magnitude depending on the system parameters. 

\subsection{Topological phase diagram and TESs}
\label{sec:phase_diagrams}

\begin{figure}
\includegraphics{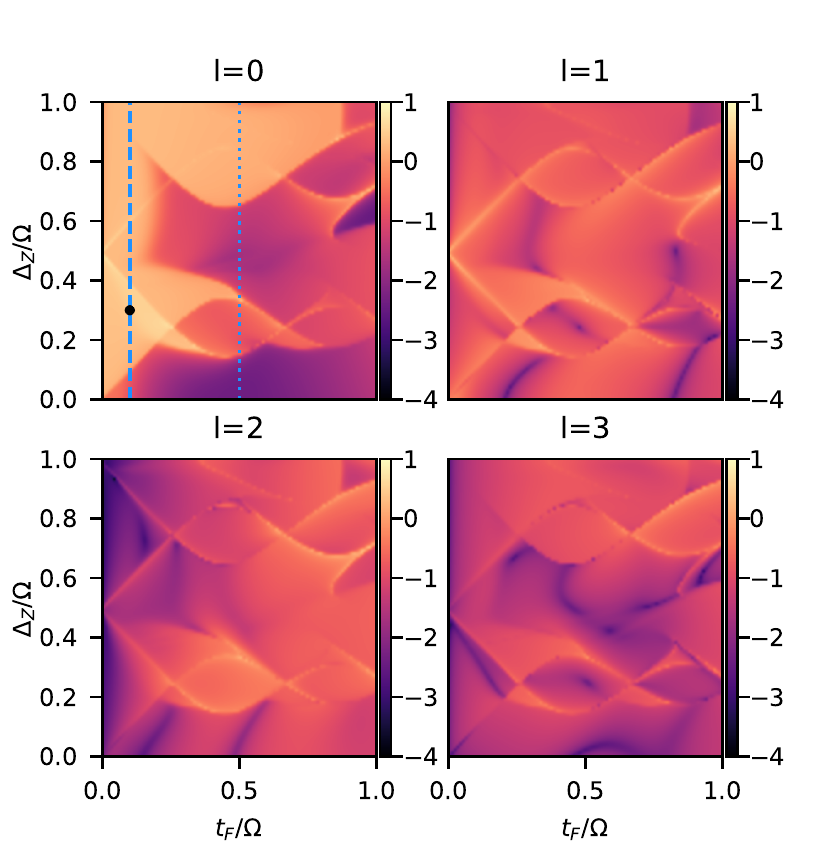}
\caption{Components of the differential conductance $\mathcal{L} g$ at $\mu_0=\frac{\Omega}{4}$ as a function of $t_F$ and $\Delta_Z$ for the case of a hybridization function which uniformly mixes between band- and spin-indices. The vertical lines in the upper left subplot indicate the values of $t_F$ ($t_F=0.1(0.5)\,\Omega$ for the dashed (dotted) line respectively) that are used in all corresponding plots of Fig.~\ref{fig:energy_resolved_tf=0.1} (dashed line) and Fig.~\ref{fig:energy_resolved_tf=0.5} (dotted line) respectively. The dot on the dashed line represents the values of $t_F$ and $\Delta_Z$ used in Fig.~\ref{fig:single_point_energy_resolved}. \label{fig:phase_diagram_mixing}}
\end{figure}

\begin{figure}
\includegraphics{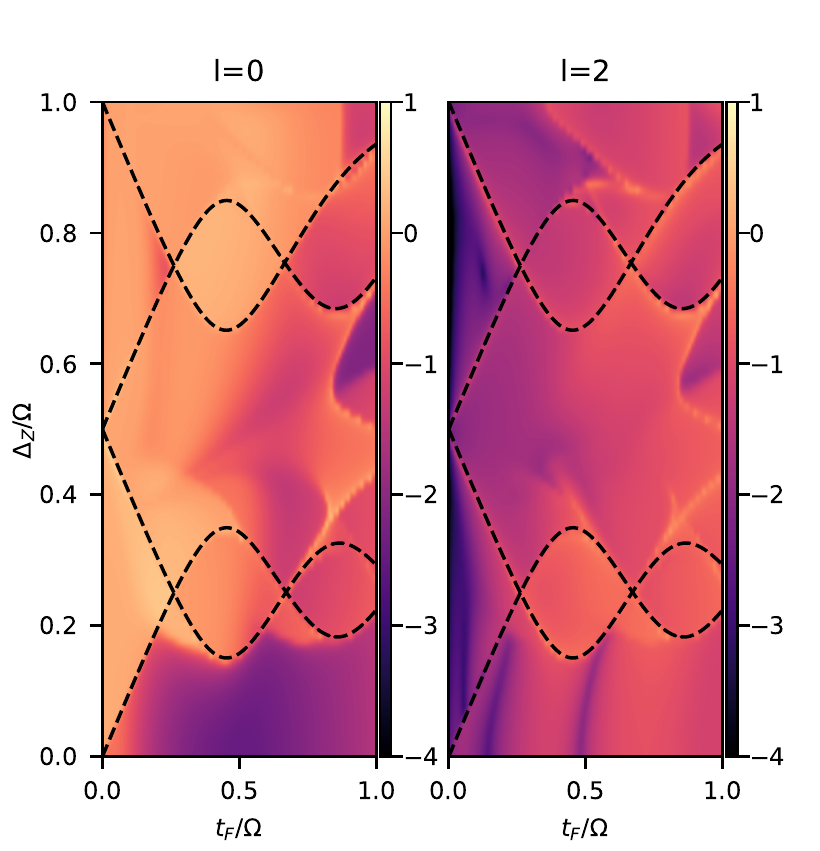}
\caption{Components of the differential conductance $\mathcal{L} g$ at $\mu_0=\frac{\Omega}{4}$ as a function of $t_F$ and $\Delta_Z$ for a hybridization function which is diagonal in band- and spin-space. Type \emph{A} phase boundaries from Fig.~\ref{fig:predicted_phase_diagram} are shown as dashed lines. \label{fig:phase_diagram_nonmixing}}
\end{figure}

\begin{figure}
\includegraphics{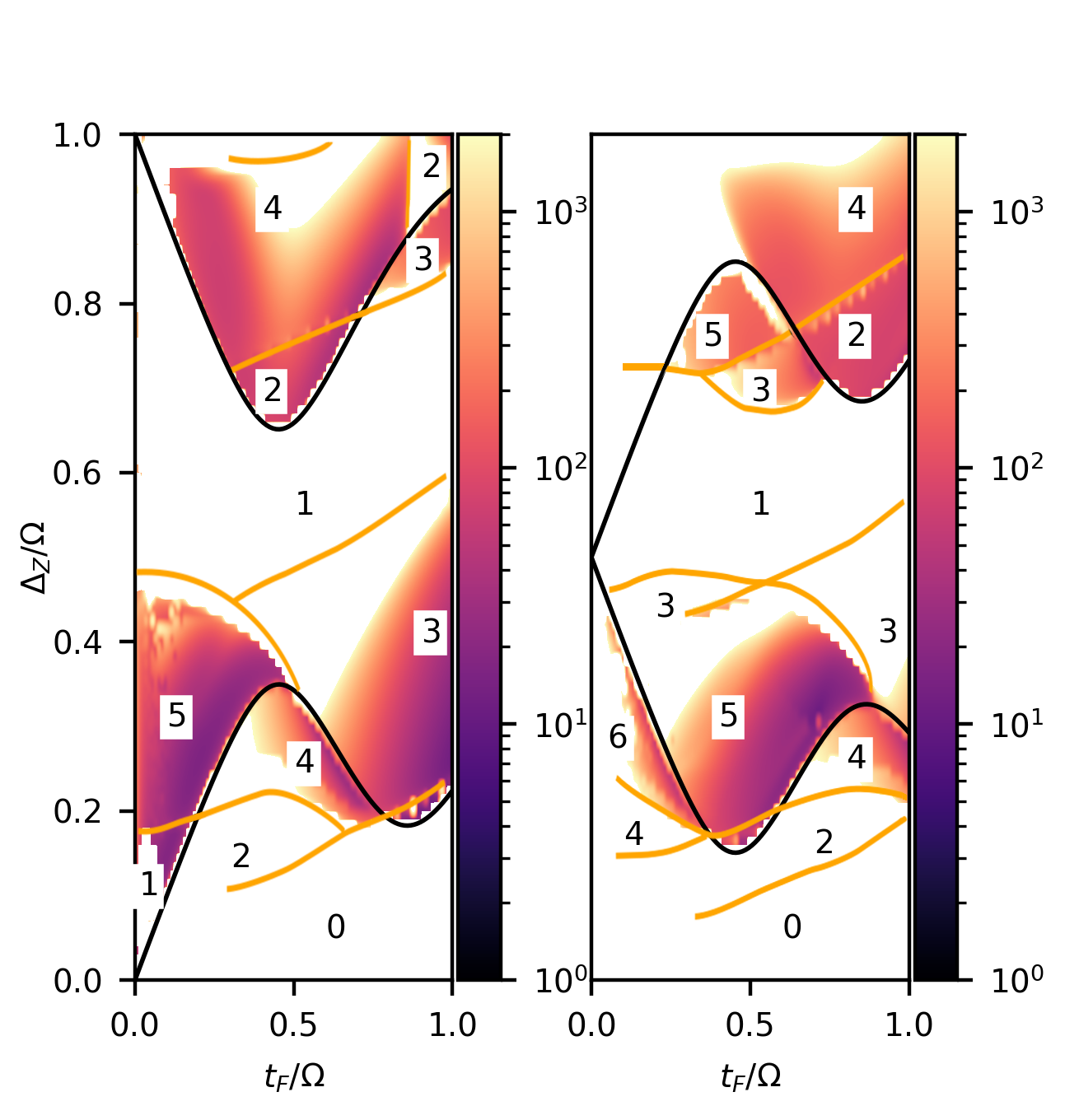}
\caption{Topological phase diagram of the system, cf. [\onlinecite{kennes_19}] for the subspace of even (left) and odd (right) Floquet indices respectively. The truncations of these correspond to a decomposition of the truncation of the full Hamiltonian chosen for the data shown in Figs.~\ref{fig:phase_diagram_mixing} and \ref{fig:phase_diagram_nonmixing}. The numbers in boxes represent the topological invariant, i.e.~the number of TESs for the given truncation. Colorbars represent the localization length of the best-localized topological edge state, where no color corresponds to either no TESs or all states having localization length larger than 2000 unit cells. Phase boundaries are denoted by black lines (here, the topological invariant changes by an odd number and the gap closing occurs at quasimomentum $k=0$) and orange lines (here, the topological invariant changes by an even number and the gap closing occurs at finite quasimomentum) respectively.  \label{fig:predicted_phase_diagram}}
\end{figure}

In order to address the TESs, the Fermi energy of the reservoirs $\mu_0$ must be tuned to the center of the topological gap that emerges in the spectrum of the truncated Floquet Hamiltonian. Choosing a symmetric truncation in the RWA basis, this corresponds to $\mu_0 = \frac{\omega}{2} = \frac{\Omega}{4}$ in the canonical basis. The effective Fermi energy entering the formula for the differential conductance Eq.~(\ref{eq:g_l}) is thus $\mu_{\eta} = 0$ and $\mu_{\eta}=\omega$ for $\eta = \pm$ respectively. 

We consider two different types of reservoirs with regard to the structure in band-/spin-space of the corresponding hybridization functions: uniform mixing of both band-/spin-indices and no mixing. There is one intrinsic difference between these two cases, which is the fact that for a diagonal hybridization function there are no terms in the effective Floquet Hamiltonian coupling even and odd Floquet indices, i.e.~$\bar{G}^R_{ll^{\prime}}=0$ if $l-l^{\prime}$ is an odd integer. From the differential conductance formula Eq.~(\ref{eq:g_l}) one can see that in this case the odd Fourier components of the differential conductance identically vanish. This is related to the fact that, without mixing, the effective Hamiltonian in the RWA basis is $\Omega$-periodic, while the mixing terms introduce an $\omega$-periodicity upon being transformed into the RWA basis. 

Varying $t_F,\Delta_Z \in (0,\Omega]$ we compute the components $\{\mathcal{L} g_l\}_{l=0,1,2,3}$ for the uniform mixing case and $\{\mathcal{L} g_l\}_{l=0,2}$ for the diagonal case and show the corresponding data in Fig.~\ref{fig:phase_diagram_mixing} and Fig.~\ref{fig:phase_diagram_nonmixing} respectively. Comparing these two figures we conclude that, besides the fact that band-mixing reservoirs induce $\omega$-periodic terms in the (otherwise $\Omega$-periodic) differential conductance, there is no qualitative difference between (a) the uniform mixing and non-mixing cases, and (b) between the various Fourier components in each of these two cases. 

Next, we compare the differential conductance data to the topological phase diagram. For technical details on the calculation of the latter the reader is referred to [\onlinecite{kennes_19}]. Note that here the aforementioned decomposition into even/odd Floquet subspaces can be performed, i.e.~the Hamiltonian can be represented as the sum of two terms projected onto the subspaces of even ($l,l^{\prime}\in\{0,\pm 2, \dots, \pm 2 \lfloor \frac{l_m}{2}\rfloor \}$) and odd ($l,l^{\prime}\in\{\pm 1, \pm 3 , \dots, \pm (2 \lfloor \frac{l_m+1}{2}\rfloor -1 )  \}$) Floquet indices respectively. The total phase diagram is then given by the sum of the two diagrams corresponding to the given odd/even truncations. Fig.~\ref{fig:predicted_phase_diagram} shows the phase diagrams corresponding to the five (left subplot) and six (right subplot) RWA-replica phase diagrams, which together constitute a truncation at $l_m=5$. Note that there are two types of phase boundaries in Fig.~\ref{fig:predicted_phase_diagram}, which correspond to phase transitions of type \emph{A} (black lines) and type \emph{B} (orange lines) respectively. Type \emph{A} phase transitions correspond to gap closings at the center of the Brillouin zone (BZ), i.e.~at $k=0$, where the change of the topological invariant is given by an odd integer, while phase transitions of type \emph{B} correspond to gap closings at finite quasimomenta, where the change of the topological invariant is given by an even integer.

{Comparing Figs.~\ref{fig:phase_diagram_mixing} and \ref{fig:phase_diagram_nonmixing} with Fig.~\ref{fig:predicted_phase_diagram} one can clearly see that all type \emph{A} phase boundaries and some of the type \emph{B} phase boundaries are reproduced in the conductance data. At these phase boundaries the magnitude of the edge-LDOS features jumps of up to five orders of magnitude, signaling a gap-closing, and, associated therewith, the divergence of the localization lengths of the responsible TESs. As expected, regions with strongly localized TESs are associated with a uniformly larger conductance compared to regions with weakly localized or no TESs. We have thus established a quantitative connection between the topological phase diagram, containing information on the number and localization lengths of TESs, and the observable transport properties of this Floquet system. Furthermore, we note that the microscopic details of the reservoirs only quantitatively influence the simulation results, demonstrating that the observed effects reflect universal physics, not setup-specific details. The two cases we consider here are of course idealized limits and in experiments we anticipate to see some intermediary behavior. In what follows we consider (for simplicity) only the uniformly mixing case. }

\subsection{Finite energy conductance and non-topological edge states}
\label{sec:ntes}

\begin{figure}
\includegraphics{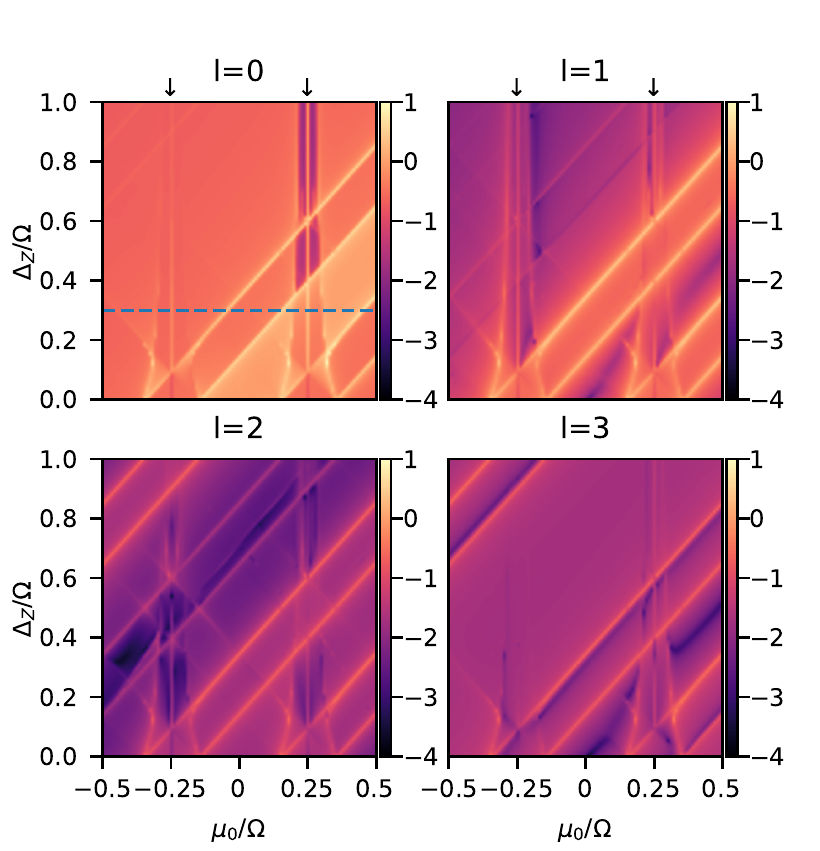}
\caption{Components of the differential conductance $\mathcal{L} g$ at $t_F=0.1\,\Omega$ as a function of $\mu_0$ and $\Delta_Z$, corresponding to the dashed vertical line in Fig.~\ref{fig:phase_diagram_mixing}. The arrows on top of the upper plots indicate the positions of the TES peaks. A cut of the data along the dashed horizontal line is shown in Fig.~\ref{fig:single_point_energy_resolved}. \label{fig:energy_resolved_tf=0.1}}
\end{figure}

\begin{figure}
\includegraphics{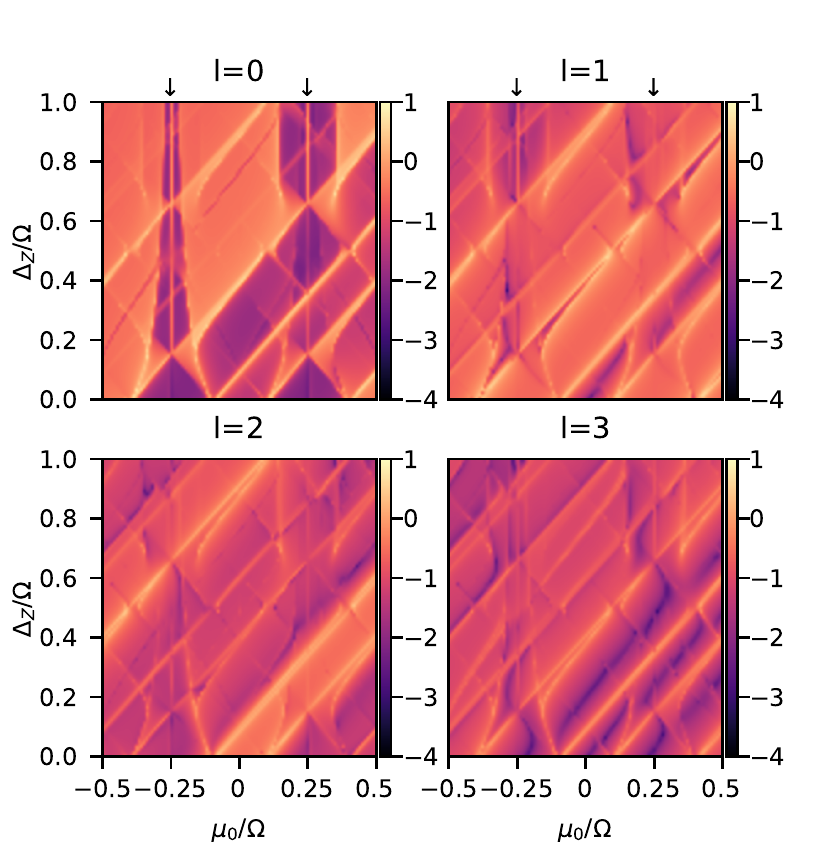}
\caption{Components of the differential conductance $\mathcal{L} g$ at $t_F=0.5\,\Omega$ as a function of $\mu_0$ and $\Delta_Z$, corresponding to the dotted vertical line in Fig.~\ref{fig:phase_diagram_mixing}. The arrows on top of the upper plots indicate the positions of the TES peaks. \label{fig:energy_resolved_tf=0.5}}
\end{figure}

\begin{figure}
\includegraphics{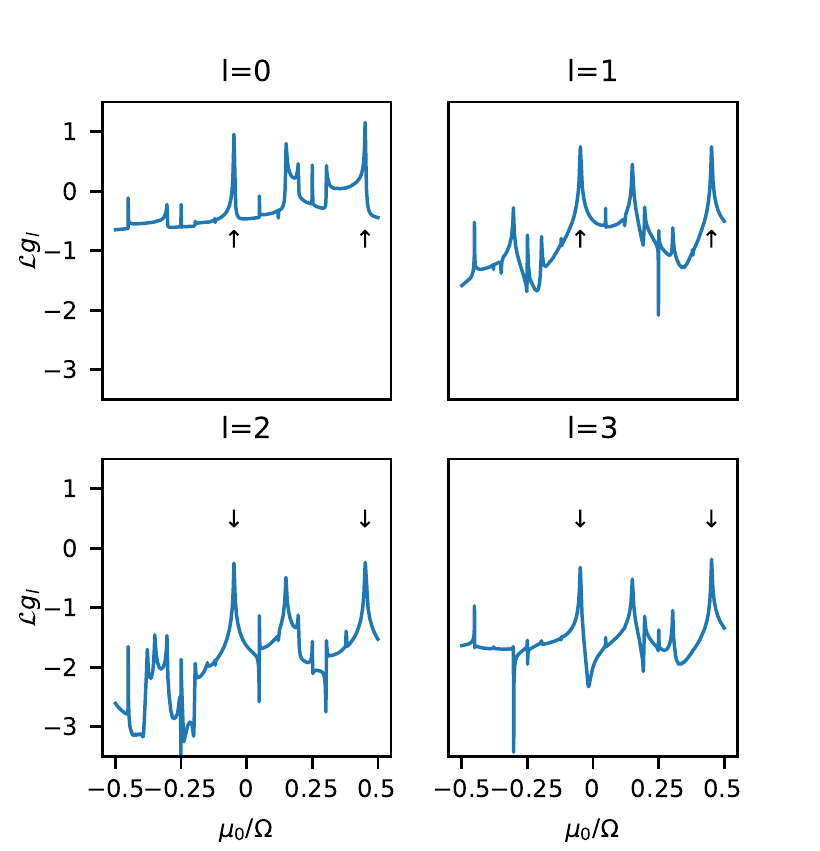}
\caption{Components of the differential conductance $\mathcal{L} g$ at $t_F=0.1\,\Omega$ and $\Delta_Z = 0.3\,\Omega$ as a function of $\mu_0$, corresponding to the horizontal line in Fig.~\ref{fig:energy_resolved_tf=0.1}. Arrows indicate some of the maxima of bands of NTESs. \label{fig:single_point_energy_resolved}}
\end{figure}

\begin{figure}
\includegraphics{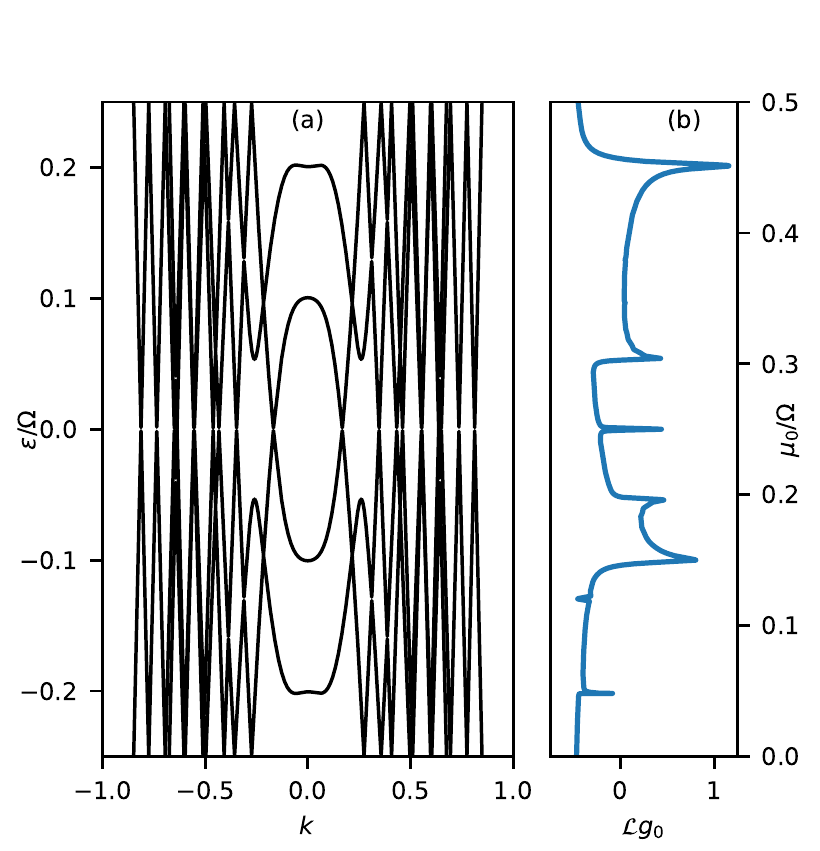}
\caption{(a) Section of the quasienergyspectrum of the Floquet Hamiltonian for $t_F=0.1\,\Omega$ and $\Delta_Z = 0.3\,\Omega$. Note the (on this scale) tiny gap around $\epsilon=0$. (b) $\mathcal{L}g_0$ component at $t_F=0.1\,\Omega$ and $\Delta_Z = 0.3\,\Omega$ on the energy interval corresponding to the spectrum shown on the left (c.f.~upper left subplot of Fig.~\ref{fig:single_point_energy_resolved}). Note that the TES peak in the conductance data occurs at $\mu_0=\frac{\Omega}{4}$, which corresponds to an effective chemical potential of $\mu_{\eta}=0,\omega$, while the topological gap of the Floquet Hamiltonian is located at $\epsilon=0$. This is the reason why there is an offset between the $y$-axes of the two subplots. Also note that, as discussed above, we do not expect the differential conductance to respect the chiral symmetry of the Hamiltonian, i.e.~the quantity $\mathcal{L}g_0(\mu_0)$ is not symmetric with respect to $\mu_0-\frac{\Omega}{4}\rightarrow -\left(\mu_0-\frac{\Omega}{4}\right)$. \label{fig:bandstructure_at_ntes}}
\end{figure}

Since TESs can only exist inside of topological gaps, tuning the Fermi energy away from the value used in the previous section, one expects to see no additional peaks with comparable height to the TES peaks. Fixing $t_F=0.1\,\Omega$ and $t_F=0.5\,\Omega$ respectively, we compute the components $\{\mathcal{L} g_l\}_{l=0,1,2,3}$ as functions of $\mu_0\in[-\frac{\Omega}{2},\frac{\Omega}{2}]$ and $\Delta_Z\in (0,\Omega]$ and show the corresponding data in Figs.~\ref{fig:energy_resolved_tf=0.1} and \ref{fig:energy_resolved_tf=0.5} respectively. 

There are several interesting features to note in these plots: At $\mu_0=\frac{\Omega}{4}$ we find a contribution from the TESs with a gap surrounding this peak (marked by arrows on top of the upper subplots). From the phase diagrams shown in the previous section we find that at $t_F=0.1(0.5)\,\Omega$ the system undergoes four phase transitions of type \emph{A} as $\Delta_Z$ is varied between $0$ and $\Omega$. These phase transitions are reproduced by the gap closings which are visible as diagonally traversing peaks intersecting the TES peak in the gap-center (i.e.~at $\mu_0=\frac{\Omega}{4}$) when $\Delta_Z$ corresponds to a phase transition point. Note, however, that not all components $\mathcal{L} g_l$ show all of the gap closings, which motivates the investigation of the full ac-conductance. 

There is also a (slightly weaker) signal from the first replica of the TESs at $\mu_0=-\frac{\Omega}{4}$ (also marked by arrows), which shows identical features, i.e.~gap closings at type \emph{A} phase transition points, a central TES peak and a gap surrounding this peak. 

Finally, the diagonally traversing peaks, which often show even stronger signals than the TES peak, must be interpreted as narrow bands of states with a large relative edge weight, living at finite energy relative to the gap center. These are what we call non-topological edge states. Cuts along the dashed line in Fig.~\ref{fig:energy_resolved_tf=0.1} are shown in Fig.~\ref{fig:single_point_energy_resolved}, clearly demonstrating the strong signal of the NTESs (strongest peaks are marked by arrows) and the fact that they do not correspond to discrete states, but rather to extended (albeit narrow) bands.
 
{Subplot (a) of Fig.~\ref{fig:bandstructure_at_ntes} shows a section of the quasienergyspectrum of the truncated Floquet Hamiltonian in momentum space at $t_F=0.1\,\Omega$ and $\Delta_Z = 0.3\,\Omega$. In subplot (b) we show the corresponding part of the conductance data $\mathcal{L}g_0$ (also shown in Fig.~\ref{fig:single_point_energy_resolved}). Note that, as discussed above, we do not expect the differential conductance to respect the chiral symmetry of the Floquet Hamiltonian, which explains why the lineshape is not symmetric with respect to $\mu_0=\frac{\Omega}{4}$. We report that the NTES peaks coincide with extremal points of bands, which occur both at $k=0$ and at finite momenta. Those NTESs which are related to extrema at finite momentum are also visible in Figs.~\ref{fig:energy_resolved_tf=0.1} and \ref{fig:energy_resolved_tf=0.5} as peaks which do not follow straight lines in $\Delta_Z-\mu_0$ space, but rather follow curved trajectories. In the following we will concentrate our discussion on those NTESs, which are related to $k=0$ extrema, since those are easier to analyze and usually show stronger signals. }

{We note that the extremal points of the dispersion hosting NTESs also inevitably correspond to van-Hove singularities (VHSs) in the bulk of the wire, which in one-dimensional systems lead to a high degeneracy of single-particle states and thus to a large bulk-LDOS. The observed conductance peaks at these energies therefore do not completely vanish in the bulk, but rather smoothly pass over into the VHS peaks, as the probe lead is moved towards the bulk. As mentioned in the introduction, the relationship of the length scales on which the exponentially localized component of the NTES decays and on which the LDOS converges to its constant, VHS-influenced bulk value, determines whether a clear separation of these two effects is possible. A study of the precise relationship between NTESs and VHSs and their relative contributions to conductance peaks at extrema of the dispersion both at the edge and in the bulk will be subject of future works [\onlinecite{muller_unpublished}]. }

\subsection{Emergence of NTESs}
\label{sec:emergence_of_ntes}

In this section we attempt to explain the emergence of NTESs in terms of a theory explaining all the properties we observed in the previous section. {red}{In Appendix~\ref{sec:ntes_rwa} we present a more detailed discussion of the easiest, and therefore analytically accessible, case of the lowest truncation in Floquet space, where only four bands are present.}

We begin by noting that at type \emph{A} phase transition points, i.e.~gap closings at quasimomentum $k=0$, the NTESs emerge at energies corresponding to the (former) gap-center. We thus suspect that their physics is related to properties of the Hamiltonian at the center of the BZ. We note that at $k=0$ the Floquet Hamiltonian takes the following form
\begin{align}
\nonumber
(\bar{h}^F_{k=0})_{ll^{\prime}} = &(\Delta_Z \sigma_x + t_F \eta_x - l\omega)\delta_{ll^{\prime}} \\
&+ t_F (\delta_{l,l^{\prime}-2} \, \eta_+ + \delta_{l,l^{\prime}+2} \, \eta_-).
\end{align}
{One can easily see that the corresponding spectrum is given by $E_i^{\pm} = \pm \Delta_Z+f_i(t_F,\omega)$, where $f_i$ are universal functions that only depend on the truncation of the Floquet Hamiltonian. Since the spectrum is symmetric with respect to $k\rightarrow -k$ every (non-flat) band features a local extremal point at $k=0$. Furthermore, at type \emph{A} phase transition points the gap closes and two of the eigenvalues $E_i^{\pm}$ vanish identically. This suggest, that NTESs emerge near these eigenvalues of the Floquet Hamiltonian at $k=0$, which also happen to be extremal points of the dispersion relation. The results shown for the special case of $t_F=0.1\,\Omega,\Delta_Z = 0.3\,\Omega$ in Fig.~\ref{fig:bandstructure_at_ntes} clearly support this claim. Note however, that the size of these peaks varies drastically from one extremal point to another, with some of the smallest peaks being invisible on the scale chosen in Fig.~\ref{fig:bandstructure_at_ntes}.   }

To better understand the connection between eigenstates of the finite-size tight-binding Hamiltonian and the physics of momentum space we consider a half-infinite system in real space with unit cells $n\in \mathbb{N}$ and vanishing boundary conditions of all components of the wavefunction on the 0th unit cell. A (delta-function) normalizable eigenstate $\psi_{\epsilon}(n)$ at energy $\epsilon$ of the corresponding Hamiltonian $h_k$ can be constructed as follows: We first determine the (complex) roots $k_i$ of the polynomial $\det\left(h_k-\epsilon\right)$ and the corresponding eigenvectors $\chi_{k_i}$ obeying $h_{k_i} \chi_{k_i} = \epsilon \chi_{k_i}$. We then construct the linear combination $\psi_{\epsilon}(n)=\sum_i \lambda_i e^{ik_i n} \chi_{k_i}$ where we only include roots with $\Im k_i \geq 0$ in order to ensure (delta)-normalizability. Finally, we impose the vanishing boundary condition at the 0th unit cell on the coefficients $\lambda_i$, i.e.~we solve the linear system of equations $\sum_i \lambda_i \chi_{k_i} = 0$. If such a set of coefficients exists, we found an eigenstate of the (real-space) half-infinite Hamiltonian at energy $\epsilon$. 

The number of boundary conditions that need to be fulfilled is equal to the dimension of the Hamiltonian $D$ (for our model we have $D=4(2l_m+1)$). The number of roots $k_i$ is given by $2D$, where, due to hermiticity, every root on the upper half-plane has a partner on the lower half-plane with equal real part, i.e.~$D=M_0 + M$ where $2M_0$ denotes the number of purely real roots (note that if $k\in\mathbb{R}$ is a root, so is $-k$) and $2M$ denotes the number of roots with finite imaginary part. The number of available roots for the construction of a normalizeable state (i.e.~roots which obey $\Im k_i\geq 0$) is thus given by $2M_0 + M = D + M_0$, which implies that there is a $M_0$-fold degeneracy at energy $\epsilon$ if this energy intersects with $M_0$ bands. 

{At energies which do not intersect with all bands (in our model this is always the case due to chiral symmetry) the degenerate subspace of eigenstates will contain at least one state with a finite contribution $e^{ik_i n} \chi_{k_i}$ with $\Im k_i>0$, i.e.~an edge state component. In general, there is no reason why the coefficient associated with this component would be of comparable magnitude to the coefficients corresponding to purely oscillating roots. This may however not be true near extrema of the bandstructure, i.e.~points where several roots meet at the origin of the complex $k$-plane (also sometimes called bifurcation points). As one approaches such a point from the side with a larger value of $M_0$ the eigenvectors $\chi_{\pm k}$ corresponding to the smallest inflowing real roots (i.e.~those which meet in the origin at the bifurcation point) gradually become colinear (linearly dependent), making it necessary to give some edge state component an increasingly larger weight in order to still fulfill the boundary condition (see Appendix~\ref{sec:ntes_rwa} for the explicit construction of such states and a discussion of the relative magnitudes of coefficients near extremal points for the analytically accessible case of just a single Floquet replica). After crossing the degeneracy point the two real roots which met at the origin start to flow out into the upper and lower half planes respectively, reducing the degeneracy by one and providing an additional edge state component. Note that this mechanism relies on the aforementioned breaking of inversion symmetry: if the system were inversion-symmetric, i.e.~$h_{-k} = h_k$, the eigenvectors $\chi_{\pm k}$ would always be colinear and nothing abrupt would happen at extremal points. Indeed, for vanishing Rashba constant $\alpha$, where inversion symmetry is restored, we report the absence of NTES peaks in the conductance data. }

{Note that at the extremal points the (purely imaginary) momenta responsible for the edge state components of the NTESs vanish, while the amplitudes associated with these components approach a maximum (see Appendix~\ref{sec:ntes_rwa} for more details). As a consequence there exist \emph{sweet-areas} in energy-space on both sides of the extremal point where, on the one hand, the localization length of the edge state component is not too large (i.e.~the imaginary part of the root is not too small), while, on the other hand, the amplitude of this component is large enough, such that the edge-LDOS is considerably enhanced relative to the bulk-average on the scale of the corresponding localization length. It is thus meaningless to speak of localization lengths of NTESs, since NTESs occur in bands with a continuous hierarchy of localization lengths (and localization length-dependent amplitudes). } 

{Apart from the special case of RWA, the Floquet model we consider here is too intricate for a detailed analytic examination of the properties near extremal points (for $l_m=5$ we have $D=44$ quasienergy bands and thus $88$ momenta).} For this reason we conclude our analysis here and leave more general discussions of NTESs (both corresponding to extremal points at $k=0$ and also finite momenta) as subject for future works [\onlinecite{muller_unpublished}]. We would like to stress however, that our discussion of the origin of NTESs in the model under consideration solely rests on the breaking of inversion symmetry, suggesting that this is the only requirement for a multi-channel 1D model to feature NTESs.

\section{Summary and conclusions}
\label{sec:summary}

In this work we proposed an experimental test for the topological properties of a one-dimensional Floquet topological insulator based on electronic transport probing the local density of states at the edge of the system. As a testbed we used a recently studied model involving Rashba spin-orbit coupling, a static Zeeman term and a coherent drive of the inter-band transition for which the topological phase diagram is known. {Using the Keldysh-Floquet formalism together with the wide-band limit and} the weak-coupling approximation we derived an expression for the differential conductance between the wire and additional leads coupled to it. These expressions only contain the effective Hamiltonian, rendering them independent of the occupation of Floquet states. In order to evaluate the differential conductance we {employed an inversion algorithm for block-tridiagonal matrices}, allowing us to compute the conductance for large system sizes (up to $10^7$ unit cells) efficiently. 

{Probing the center of the topological gap we reproduced the topological phase diagram via the conductance data as expected, thus demonstrating that transport experiments allow for a verification of topological properties in one-dimensional Floquet topological insulators. Surprisingly, we found unexpected peaks in the differential conductance at special energies corresponding to the bulk of the system, hinting at the presence of narrow bands of states centered around extremal points of the dispersion relation, which feature a large relative weight at the edge of the system. These states are linear combinations of delocalized bulk states and exponentially localized edge states, where the amplitude of the edge state component is sharply peaked at aforementioned extremal points. We explained the emergence of these \emph{non-topological edge states} in terms of an intuitive yet quantitative physical picture. Since this picture is not specific for the model we consider here, but only relies on a broken inversion-symmetry, we reason that the emergence of such states should be a more general phenomenon, which may be present in a larger class of systems. }

\section*{Acknowledgments}
This work was supported by the Deutsche Forschungsgemeinschaft via RTG 1995. 
Simulations were performed with computing resourced granted by RWTH Aachen University under projects rwth0347, rwth0362 and rwth0473.

\begin{appendix}

\section{Keldysh-Floquet formalism}
\label{sec:keldysh-floquet_formalism}

\begin{figure}
\includegraphics{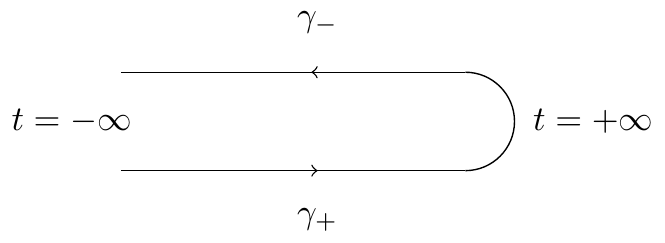}
\caption{Diagram depicting the Keldysh-contour, which runs from $-\infty$ to $+\infty$ on the lower branch ($\gamma_+$) and then back to $-\infty$ on the upper branch ($\gamma_-$). \label{fig:keldysh_contour}}
\end{figure}
In this section we review some details of the Keldysh-Floquet formalism which are necessary for the computation of the transport properties. For more detailed references the reader is referred to e.g.~[\onlinecite{eissing_16_prb}-\onlinecite{liu_17}]. 
\subsection{Contour-ordered Green's function and Dyson equation}
The central object of interest in Keldysh formalism is the contour-ordered Green's function
\begin{equation}
\label{eq:A1.1}
G_{\alpha \alpha^{\prime}}(t,t^{\prime}) = -i \left< T_{\gamma} c_{\alpha}(t) c_{\alpha^{\prime}}^{\dagger}(t^{\prime})\right>.
\end{equation}
Here, the time arguments $t,t^{\prime}$ do not live on the real number line, but rather on the Keldysh contour, which is depicted in Fig.~\ref{fig:keldysh_contour}. The operators $c_{\alpha}(t)/c^{\dagger}_{\alpha}(t)$ annihilate/create a fermion in the single-particle state $\alpha$ at time $t$. $T_{\gamma}$ denotes the contour-ordering operator, which orders the succeeding operators with respect to the position of their time-arguments on the Keldysh contour, taking into account anti-commutation for fermionic operators and $\left<\bullet\right>$ denotes averaging with respect to the initial (equilibrium) density matrix. 

The two time arguments $t,t^{\prime}$ of the contour-ordered Green's function can live on either of the two branches of the Keldysh contour, defining four distinct but related Green's functions:
\begin{align}
\label{eq:A1.2}
&t\in \gamma_+ , t^{\prime} \in \gamma_+ : \quad G^c_{\alpha\alpha^{\prime}}(t,t^{\prime}) = -i \left< T c_{\alpha}(t) c_{\alpha^{\prime}}^{\dagger}(t^{\prime})\right>, \\
\label{eq:A1.3}
&t\in \gamma_+ , t^{\prime} \in \gamma_- : \quad G^<_{\alpha\alpha^{\prime}}(t,t^{\prime}) = i \left<c_{\alpha^{\prime}}^{\dagger}(t^{\prime}) c_{\alpha}(t) \right>, \\
\label{eq:A1.4}
&t\in \gamma_- , t^{\prime} \in \gamma_+ : \quad G^>_{\alpha\alpha^{\prime}}(t,t^{\prime}) = -i \left< c_{\alpha}(t) c_{\alpha^{\prime}}^{\dagger}(t^{\prime})\right>, \\
\label{eq:A1.5}
&t\in \gamma_- , t^{\prime} \in \gamma_- : \quad G^{\tilde{c}}_{\alpha\alpha^{\prime}}(t,t^{\prime}) = -i \left< \tilde{T} c_{\alpha}(t) c_{\alpha^{\prime}}^{\dagger}(t^{\prime})\right> ,
\end{align} 
where $T/\tilde{T}$ denote the usual time-/anti-time-ordering operators. The objects in Eqs.~(\ref{eq:A1.2})-(\ref{eq:A1.5}) are known as the chronological, lesser, greater and anti-chronological Green's functions. In addition, we define:
\begin{align}
\label{eq:A1.6}
G^R_{\alpha\alpha^{\prime}}(t,t^{\prime}) &= -i \Theta(t-t^{\prime}) \left< \left\{ c_{\alpha}(t),c^{\dagger}_{\alpha^{\prime}}(t^{\prime})\right\}\right>, \\
\label{eq:A1.7}
G^A_{\alpha\alpha^{\prime}}(t,t^{\prime}) &= i \Theta(t^{\prime}-t) \left< \left\{ c_{\alpha}(t),c^{\dagger}_{\alpha^{\prime}}(t^{\prime})\right\}\right>, \\
\label{eq:A1.8}
G^K_{\alpha\alpha^{\prime}}(t,t^{\prime}) &= -i \left< \left[ c_{\alpha}(t),c^{\dagger}_{\alpha^{\prime}}(t^{\prime})\right]\right>, 
\end{align}
which are known as the retarded, advanced and Keldysh Green's functions. Note the following useful relations:
\begin{align}
\label{eq:A1.9}
\left(G^>-G^<\right)_{\alpha\alpha^{\prime}}(t,t^{\prime}) &= \left(G^R-G^A\right)_{\alpha\alpha^{\prime}}(t,t^{\prime}),  \\
\label{eq:A1.10}
G^A(t^{\prime},t) &= \left(G^R(t,t^{\prime})\right)^{\dagger}.
\end{align}
Using a diagrammatic expansion in terms of the self-energy $\Sigma$ one can derive the Dyson equation for the retarded Green's function 
\begin{equation}
\label{eq:A1.11}
\left[i\partial_t - h(t)\right] G^R(t,t^{\prime}) = \delta(t-t^{\prime}) + \int \mathrm{d}t_1 \Sigma^R(t,t_1) G^R(t_1,t^{\prime})
\end{equation}
with the single-particle Hamiltonian $h(t)$ and the retarded self-energy $\Sigma^R$. Using Eqs.~(\ref{eq:A1.10}) and (\ref{eq:A1.11}) and some approximation to $\Sigma^R$, one can compute $G^{R/A}$. The greater and lesser Green's functions can then be computed as follows
\begin{equation}
\label{eq:A1.12}
G^{\gtrless}(t,t^{\prime}) = \int \int \mathrm{d}t_1 \mathrm{d}t_2 G^R(t,t_1) \Sigma^{\gtrless}(t_1,t_2) G^A(t_2,t^{\prime}).
\end{equation}
Finally, in order to compute the various components of the self-energy, we need to give explicit expressions for the corresponding free Green's functions $g$ for a static system:
\begin{align}
\label{eq:A1.13}
g^R(t,t^{\prime}) &= \int \frac{\mathrm{d}E}{2\pi} \frac{e^{-iE(t-t^{\prime})} }{E-h+i0^+}, \\ 
\label{eq:A1.14}
g^<(t,t^{\prime}) &= \int \frac{\mathrm{d}E}{2\pi} e^{-iE(t-t^{\prime})} 2\pi i \delta(E-h) n_F(E), \\ 
\label{eq:A1.15}
g^>(t,t^{\prime}) &= \int \frac{\mathrm{d}E}{2\pi} e^{-iE(t-t^{\prime})} 2\pi i \delta(E-h) \left[n_F(E)-1\right],
\end{align}
where $h$ denotes the static single-particle Hamiltonian and $n_F(E)$ denotes the Fermi function.
\subsection{Floquet Green's functions} 
We will now solve the Dyson equation for the retarded Green's function for a time-periodic single-particle Hamiltonian $h(t)=h(t+T)$ with period $T=\frac{2\pi}{\omega}$. In order to keep the problem as simple as possible we assume a time-local self-energy $\Sigma^R(t,t^{\prime}) = \Sigma^R \delta(t-t^{\prime})$. Using this restriction the Dyson equation (see Eq.~(\ref{eq:A1.11})) reads
\begin{equation}
\label{eq:A1.16}
\left[i\partial_t - h(t) - \Sigma^R \right] G^R(t,t^{\prime}) = \delta(t-t^{\prime}).
\end{equation}
Defining the Floquet Hamiltonian and Floquet Green's function as
\begin{align}
\label{eq:A1.17}
h^F_{ll^{\prime}} &= \int\displaylimits_0^T\frac{\mathrm{d}t}{T}e^{i(l-l^{\prime})\omega t} h(t) - l\omega \delta_{ll^{\prime}}, \\
G^R_{ll^{\prime}}(E) &\equiv G^R_{l-l^{\prime}}(E+l^{\prime}\omega), \\
G^R_l(E) &= \int\displaylimits_0^T\frac{\mathrm{d}t}{T}e^{il\omega t} \int \mathrm{d}t^{\prime} e^{iE(t-t^{\prime})} G^R(t,t^{\prime})
\end{align} 
we can rewrite Eq.~(\ref{eq:A1.16}) as follows
\begin{equation}
\label{eq:A1.18}
\sum_{l_1} (E \delta_{ll_1}-h^F_{ll_1}-\Sigma^R\delta_{ll_1}) G^R_{l_1l^{\prime}}(E) = \delta_{ll^{\prime}},
\end{equation}
whose solution reads
\begin{equation}
\label{eq:A1.19}
G^R_{ll^{\prime}}(E) = \left(\frac{1}{E-h^F-\Sigma^R}\right)_{ll^{\prime}}.
\end{equation}
Having inverted the matrix in Eq.~(\ref{eq:A1.19}) we can then express the real-time retarded Green's function as 
\begin{equation}
\label{eq:A1.22}
G^R(t,t^{\prime}) = \sum_l e^{-il\omega t} \int \frac{\mathrm{d}E}{2\pi} e^{-iE(t-t^{\prime})} G^R_{l0}(E).
\end{equation}

\section{Estimate of corrections beyond weak-coupling approximation}
\label{sec:beyond_WCA}
Here, we give a derivation of the corrections to the differential conductance formula beyond the weak-coupling approximation (WCA) and estimate its influence on the physical results. 

Using Eqs.~(\ref{eq:A1.12}) and (\ref{eq:5a}) we find
\begin{align}
\nonumber
\partial_V G^{\gtrless}(t,t^{\prime})\rvert_{V\rightarrow 0} = \frac{ie}{2\pi} \int \int \mathrm{d}t_1 \mathrm{d}t_2 \times \\
\left\{e^{-i\mu_0(t_1-t_2)} G^R(t,t_1) \Gamma_p G^A(t_2,t^{\prime}) \right\}.
\end{align}
Using Eqs.~(\ref{eq:5}) and (\ref{eq:6}) we arrive at the identity 
\begin{equation}
\Sigma^>(t,t^{\prime})-\Sigma^<(t,t^{\prime}) = -i \Gamma \delta(t-t^{\prime}).
\end{equation}
Denoting the corrections to the differential conductance beyond WCA by $g_{bWCA}(t)$ we find
\begin{align}
\nonumber
g_{bWCA}(t) = -\frac{1}{2} \mathrm{Tr} &\left\{\int \mathrm{d}t_1 e^{i\mu_0(t-t_1)} \Gamma_p G^R(t,t_1)\right\} \\
\times& \left\{\int \mathrm{d}t_2 e^{i\mu_0(t_2-t)} \Gamma_p G^A(t_2,t)\right\}
\end{align}
We thus conclude that this correction is at least of order $\lesssim \mathcal{O}\left(\frac{\Gamma_p}{\Gamma_s}\right)^2$ and thus negligible under the assumptions of the WCA. Furthermore, we note that the correction only depends on expressions which also appear in the formula given in the main text (i.e.~the Fourier transform of the retarded Green's function), which suggests that no new features can appear in the correction beyond WCA. A crucial difference however lies in the fact that this expression, even for hybridization functions which are diagonal in position space, depends on all off-diagonal components of the Green's functions, rendering the algorithm presented in the main text useless.

\section{Non-topological edge states in the half-infinite system in rotating wave approximation}
\label{sec:ntes_rwa}
Here, we construct and analyze the non-topological edge states (NTESs) of a half-infinite wire using the smallest truncation in Floquet space, also known as rotating wave approximation (RWA).

The RWA momentum-space Hamiltonian is given by
\begin{equation}
h^R_k = (E_k + \alpha_k \sigma_z)\, \eta_z + \Delta_Z \sigma_x + t_F \eta_x,
\end{equation}
where $E_k = W\sin^2 \frac{k}{2}$ and $\alpha_k = \alpha \sin k$ with quasimomentum $k$. We transform this Hamiltonian into the chiral basis, which is spanned by the eigenstates of the chiral symmetry operator $\mathcal{S}=\eta_y\sigma_z$ and find that the Hamiltonian assumes a block-off-diagonal structure in this basis ($h^R_k = \begin{pmatrix} 0 & A_k \\ A_k^{\dagger} & 0 \end{pmatrix}$), where we have defined
\begin{equation}
A_k = \Delta_Z + E_k \tau_x + (t_F+i\alpha_k )\tau_y
\end{equation}
with Pauli matrices $\tau_i$. In the chiral basis, the eigenvalue equation $h_k^R \chi_k = \epsilon_k \chi_k$ for Bloch states $\chi_k = \left({\chi_k^+}^T,{\chi_k^-}^T\right)^T$ can be expressed in terms of the chiral components $\chi_k^{\tau}$ as follows:
\begin{equation}
\begin{pmatrix}
A_k \chi_k^-  \\
A^{\dagger}_k \chi_k^+  
\end{pmatrix}
= \epsilon_k
\begin{pmatrix}
 \chi_k^+ \\
\chi_k^- 
\end{pmatrix}
\Rightarrow 
\begin{pmatrix}
A^{\dagger}_k A_k \chi_k^-  \\
A_k A^{\dagger}_k \chi_k^+  
\end{pmatrix} 
= \epsilon_k^2
\begin{pmatrix}
\chi_k^- \\
\chi_k^+ 
\end{pmatrix} .
\end{equation}
Defining the two vectors ($\tau=\pm $)
\begin{equation}
\xi_k^{\tau} = |\xi_k^{\tau}| n_k^{\tau} = \left(\Delta_Z E_k,\Delta_Z t_F,\tau \alpha_k E_k \right)^T
\end{equation}
we find
\begin{align}
A_k A^{\dagger}_k &= E_k^2 + \alpha_k^2 + \Delta_Z^2 + t_F^2 + 2 \xi_k^+ \cdot \tau, \\
A^{\dagger}_k A_k &= E_k^2 + \alpha_k^2 + \Delta_Z^2 + t_F^2 + 2 \xi_k^- \cdot \tau, \\
\epsilon_k^2 &= E_k^2 + \alpha_k^2 + \Delta_Z^2 + t_F^2 \pm 2 |\xi_k^{\pm}|,
\end{align}
which implies that the chiral components of the Bloch states need to fulfill the following equations:
\begin{equation}
\left(n_k^{\tau} \cdot \tau \right)\chi_k^{\tau} = \pm \chi_k^{\tau},
\end{equation}
i.e.~depending on the branch of the dispersion they correspond to (pseudo-)spin up/down along the direction given by the unit-vector $n_k^{\tau}$. In terms of spherical coordinates these vectors read
\begin{equation}
n_k^{\tau} = \left(\sin \Theta_k^{\tau} \cos \phi_k^{\tau}, \sin \Theta_k^{\tau} \sin \phi_k^{\tau} , \cos \Theta_k^{\tau} \right)^T,
\end{equation}
where
\begin{align}
\Theta_k^{\tau} &= \tau \Theta_k + \frac{1-\tau}{2} \pi, \\
\label{eq:rwa_ntes_1}
\Theta_k &= \arctan\left[\frac{\Delta_Z}{\alpha_k} \sqrt{1 + \left(\frac{t_F}{E_k}\right)^2}\right] = -\Theta_{-k} ,\\
\label{eq:rwa_ntes_2}
\phi_k^{\tau} &= \phi_k = \arctan \frac{t_F}{E_k} = \phi_{-k} .
\end{align}
For the full Bloch states we find (up to normalization):
\begin{align}
\nonumber
\chi_k^{(+)} &= \left(\cos \frac{\Theta_k^+}{2}, e^{i\phi_k^+} \sin \frac{\Theta_k^+}{2},\cos \frac{\Theta_k^-}{2},e^{i\phi_k^-} \sin\frac{\Theta_k^-}{2}\right)^T \\
&= \left(\cos \frac{\Theta_k}{2}, e^{i\phi_k} \sin \frac{\Theta_k}{2},\sin \frac{\Theta_k}{2},e^{i\phi_k} \cos\frac{\Theta_k}{2}\right)^T, \\
\nonumber
\chi_k^{(-)} &= \left(\sin \frac{\Theta_k^+}{2},- e^{i\phi_k^+} \cos \frac{\Theta_k^+}{2},\sin \frac{\Theta_k^-}{2}, - e^{i\phi_k^-} \cos \frac{\Theta_k^-}{2}\right)^T \\
&= \left(\sin \frac{\Theta_k}{2},- e^{i\phi_k} \cos \frac{\Theta_k}{2},\cos \frac{\Theta_k}{2}, - e^{i\phi_k} \sin \frac{\Theta_k}{2}\right)^T, 
\end{align}
for the $\pm$-branches of the square of the dispersion respectively. Note that one can construct the remaining two states (corresponding to branches of the dispersion at negative energies) by applying the chiral symmetry operator (in the chiral basis) to the two states above, which just changes the relative sign between the first and last two components. 

In order to construct solutions of the half-infinite system we superimpose solutions of the infinite system (i.e.~Bloch states $e^{ikn} \chi_k$) in such a way, that the open boundary conditions on an artificial $0$-th unit cell are fulfilled, while making sure that the wavefunction remains normalizable (for more details on this construction see e.g.~[\onlinecite{kennes_19}]). Fixing an energy $\epsilon$ we first determine all the allowed (complex) quasimomenta, which are solutions of the equation $\epsilon_k = \epsilon$ for a given branch of the dispersion $\epsilon_k$ (note that there are 8 such momenta for each branch, which, due to hermiticity of the Hamiltonian, are symmetrically arranged with respect to the real axis) and then try to find a set of coefficients $\left\{\lambda_i\right\}$ such that $\sum_i \lambda_i \chi_{k_i} = 0$. The full semi-infinite wavefunction is then given by $\psi_{\epsilon}(n) = \sum_i \lambda_i e^{ik_i n} \chi_{k_i}$ for $n\geq 1$, where we only include momenta with non-negative imaginary parts in order to ensure (delta-)normalizability.

\begin{figure}
\includegraphics{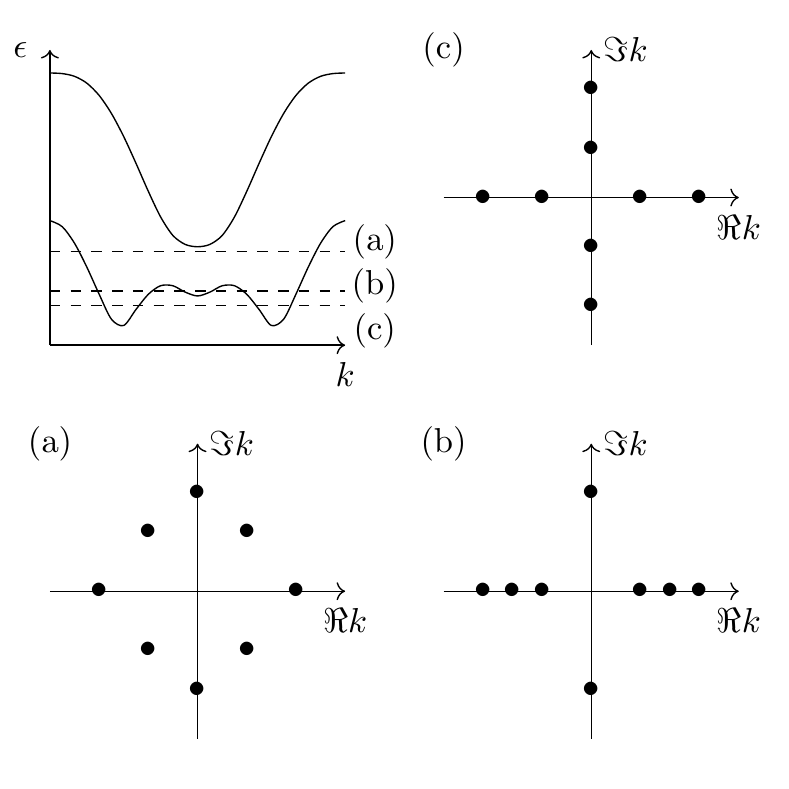}
\caption{Diagrams showing all possible scenarios of distributions of quasimomenta $k$ (solutions of the equation $\epsilon=\epsilon_k$) over the complex plane (indicated by black dots). In the upper left subplot the two positive-valued bands of the RWA Hamiltonian are sketched, with the x-axis coinciding with $\epsilon=0$. At three different energies $\epsilon$ (indicated by dashed horizontal lines) the roots of $\epsilon=\epsilon_k$ are shown in the corresponding subplots.  \label{fig:root_distribution}}
\end{figure}

In general, independent of the branch of the dispersion, the boundary conditions reduce to the following four equations:
\begin{align}
\label{eq:rwa_ntes_3}
&\sum_i \lambda_i\cos \frac{\Theta_{k_i}}{2} = 0, \\
&\sum_i \lambda_i\sin \frac{\Theta_{k_i}}{2} = 0, \\
&\sum_i \lambda_i e^{i \phi_{k_i}}\cos \frac{\Theta_{k_i}}{2} = 0,\\
\label{eq:rwa_ntes_4}
&\sum_i \lambda_i e^{i \phi_{k_i}}\sin \frac{\Theta_{k_i}}{2} = 0.
\end{align}
All possible scenarios of distributions of momenta over the complex plane are depicted in Fig.~\ref{fig:root_distribution}. In each of these cases there are at least five momenta with non-negative imaginary part and at least two of them form a pair of opposite, real momenta $\pm k$. One can easily check that four momenta, two of which form such a pair, are not sufficient to fulfill the boundary conditions. Including a fifth momentum however always suffices to fulfill these conditions (four equations and five unknowns). Before we go on to analyze the relation between the coefficients $\lambda_i$ corresponding to localized and delocalized components respectively we note, that the need to include complex momenta at all arises due to the fact that $\Theta_k$ is an odd function of $k$ (which can be traced back to the odd Rashba term $\alpha_k$). Replacing $\alpha_k$ by an even function of $k$ renders two of the four boundary conditions equivalent for a pair of plane waves with momenta $\pm k$. Breaking of the inversion symmetry due to the Rashba term thus necessitates the presence of exponentially localized components in the wavefunction and is thus vital in the emergence of NTESs.

Let us assume the case of four real momenta ($\pm k_{1/2}$) and one purely complex momentum ($i\kappa$) with coefficients $\lambda_{1/2}^{\pm}$ and $\lambda_e$ respectively. Note that this scenario covers both case (b) and case (c) as indicated in Fig.~\ref{fig:root_distribution}. The case of two real and three complex momenta (case (a)) leads to much longer expressions and is thus only discussed qualitatively in the following (relying on numerical results).

We solve the system of equations (\ref{eq:rwa_ntes_3}) - (\ref{eq:rwa_ntes_4}) for the ratios of bulk- to edge-amplitudes and find
\begin{equation}
\label{eq:rwa_ntes_5}
\frac{\lambda_{1/2}^{\pm}}{\lambda_{e}} = \frac{e^{i\phi_{{2/1}}}-e^{i\phi_{i\kappa}}}{e^{i\phi_{{1/2}}}-e^{i\phi_{{2/1}}}} \frac{\sin \frac{\Theta_{{1/2}}\pm \Theta_{i\kappa}}{2}}{\sin \Theta_{{1/2}}},
\end{equation}
where we use the convenient notation $X_{1/2}=X_{k_{1/2}}$ for $X=\Theta,\phi$. We note that, in principle, every bulk state of this model is a \emph{non-topological edge state}, since all states contain exponentially localized components. At most energies however, the amplitudes associated with these components are rather small and thus these states are almost indistinguishable from trivial bulk states. In order to find those energies where a significant portion of the NTES is actually localized at the edge of the system, we analyze the sum of the squares of the absolute values of the ratios defined in Eq.~(\ref{eq:rwa_ntes_5}), which evaluates to
\begin{align}
\nonumber
&R \equiv \sum_{\genfrac{}{}{0pt}{}{i=1,2}{\sigma=\pm}} \left| \frac{ \lambda_i^{\sigma}}{\lambda_e}\right|^2 = \\
&\frac{\left[\frac{\sin^2 \frac{\phi_2-\phi_{i\kappa}}{2}}{\sin^2 \Theta_1} \left(\cosh \Im \Theta_{i\kappa} - \cos \Re \Theta_{i\kappa} \cos \Theta_1\right) + \left(1\leftrightarrow 2\right) \right]}{\sin^2 \frac{\phi_1-\phi_2}{2}} .
\end{align}
Minima of this expression hint at energies at which the local density of states at the edge of the system (on a scale $\frac{1}{\kappa}$) is considerably larger than the bulk-average, leading to (in principle) observable NTESs. Let us analyze under which conditions this occurs.

We begin our analysis by noting that, due to the fact that $\phi_k$ is a monotonous function of $k$, the overall prefactor $\sin\left(\frac{\phi_1-\phi_2}{2}\right)^{-2}$ forces us to consider momenta $k_{1/2}$ which are reasonably far apart. Combining this fact with the observation that $\phi_{i\kappa}<0$ while $\phi_{1/2}>0$ excludes the possibility of minimizing $R$ via the terms $\sin\left(\frac{\phi_{1/2}-\phi_{i\kappa}}{2}\right)^2$. Turning our attention to the parameter $\Theta_{i\kappa}$ we note that its real part is a piece-wise discontinuous function
\begin{equation}
\Re \Theta_{i\kappa} = 
\begin{cases}
0 \quad \text{for} \quad \kappa \geq \kappa_{c} \\
-\frac{\pi}{2} \quad \text{for} \quad \kappa < \kappa_{c} \\
\end{cases},
\end{equation}
where the critical inverse localization length $\kappa_c$ is determined by the following equation
\begin{equation}
\frac{\Delta_Z}{\alpha} \frac{1}{\sinh \kappa_c} \sqrt{1+\left(\frac{t_F}{W}\right)^2 \frac{1}{\sinh^4 \frac{\kappa_c}{2}}} = 1.
\end{equation}
This critical inverse localization length is furthermore associated with a divergence of $\Im \Theta_{i\kappa}$, forcing us to consider either very small or very large $\kappa$. Note that the energy window in which there are four or more real momenta (i.e.~where this discussion is valid) is rather narrow (as can be seen from the exemplary bandstructure plot in Fig.~\ref{fig:root_distribution}), limiting the maximal value of $\kappa$ that can be achieved. The only way to consistently minimize $R$ is thus by going to the limit of very small $\kappa$, i.e.~to consider an energy close to a band minimum/maximum at $k=0$. This scenario corresponds to a bifurcation point of the dispersion, where the four solutions $\pm i\kappa, \pm k_1$ flow into the origin, with an additional band intersecting the energy $\epsilon$ typically at rather large quasimomentum $k_2\lessapprox \pi$.

We note that this discussion can in principle be repeated for the case of two real momenta ($\pm k_0$) and three complex momenta ($\pm k_1+i\kappa_1,i\kappa_2$) (case (a)). The corresponding quantity $R$, as mentioned above, is much longer and very messy so we decided to not include it here. We report however that it is minimized near energies which correspond to a bifurcation point where $k_1\pm i\kappa_1$ and$-k_1\pm i\kappa_1$ respectively meet on the real axis as $\kappa_1\to 0$, marking the transition between the case of two real momenta and six real momenta.

Let us analyze the quantity $R$ in the vicinity of a $k=0$ extremal point, i.e.~let us expand it around $k_1,\kappa\to 0$ and $k_2\to \pi$. Note the following identities:
\begin{align}
\Theta_k &\to 
\begin{cases}
\frac{\pi}{2} - \frac{\alpha W}{4\Delta_Z t_F} k^3 + \mathcal{O}(k^5)  \\
\frac{\pi}{2} - \frac{\alpha}{\Delta_Z} \frac{\pi -k }{\sqrt{1+\left(\frac{t_F}{W}\right)^2}} + \mathcal{O}(\left(\pi -k\right)^3) 
\end{cases}, \\
\phi_k &\to
\begin{cases}
\frac{\pi}{2} - \frac{W}{4t_F} k^2 + \mathcal{O}(k^4)  \\
\arctan\left(\frac{t_F}{W}\right) + \frac{1}{4} \frac{\frac{t_F}{W}\left(\pi-k\right)^2}{1+\left(\frac{t_F}{W}\right)^2}  + \mathcal{O}(\left(\pi-k\right)^4)
\end{cases}, \\
\Im \Theta_{i\kappa}&\to  - \frac{\alpha W}{4\Delta_Z t_F} \kappa^3 + \mathcal{O}(\kappa^5), \\
\phi_{i\kappa}&\to -\frac{\pi}{2} + \frac{W}{4t_F} \kappa^2 + \mathcal{O}(\kappa^4) .
\end{align}
Using these we find
\begin{equation}
R \to R_0 + l_1^2\left(R_0 k_1^2 - \kappa^2\right) + \left(l_2\left(\pi-k_2\right)\right)^2  + \mathcal{O}\left(\dots\right)^4
\end{equation}
where we have defined
\begin{align}
\label{eq:A3_1}
R_0 &= \frac{1+\sin^2 \left(\frac{\frac{\pi}{2}+\arctan\frac{t_F}{W}}{2}\right)}{\sin^2 \left( \frac{\frac{\pi}{2}-\arctan\frac{t_F}{W}}{2}\right)} = 3 + 4 \frac{t_F}{W} + \mathcal{O}\left(\frac{t_F}{W}\right)^2 ,\\
\nonumber
l_1 &= \sqrt{\frac{1}{8} \frac{W}{t_F}\frac{\cos \left(\arctan \frac{t_F}{W}\right)}{\sin^2 \frac{\frac{\pi}{2}-\arctan\frac{t_F}{W}}{2}} } \\
&= \frac{1}{2} \sqrt{\frac{W}{t_F}} + \frac{1}{4} \sqrt{\frac{t_F}{W}} + \mathcal{O}\left(\frac{t_F}{W}\right)^{\frac{3}{2}},\\
\nonumber
l_2 &= l_1 \sqrt{\frac{1+R_0 + 8\left(\frac{\alpha}{\Delta_Z} \right)^2\frac{W}{t_F} \frac{1}{\cos \arctan \frac{t_F}{W}}}{1+\left(\frac{W}{t_F}\right)^2}} \\
\label{eq:A3_2}
&= \sqrt{2} \frac{\alpha}{\Delta_Z} + \mathcal{O}\left(\frac{t_F}{W}\right) .
\end{align}
Note that although it may seem as if the various momenta $k_{1/2}$ and $\kappa$ were varied independently, because they are all roots of the eigenvalue equation $\epsilon_k=\epsilon$ for some energy $\epsilon$, they are all related and of course also depend on the system parameters in a highly nonlinear fashion. The construction presented in this section must therefore be taken with some care. It is for example not possible to infer information about the dependence of the localization length of NTESs, or the edge-LDOS on the system parameters from the results presented in Eqs.~(\ref{eq:A3_1})-(\ref{eq:A3_2}), because $k_{1/2}$ and $\kappa$ also implicitly depend on those parameters. Nonetheless, the calculation presented above implies that close to extremal/bifurcation points the relative weight of the exponentially localized components of the bulk states is peaked, which supplements and, to some degree, explains the numerical findings presented in the main text. 

Let us summarize the results of this section: Due to the inversion-symmetry breaking Rashba term, bulk states of the model under consideration are forced to contain exponentially localized components instead of just the usual oscillating, delocalized components. Near extremal points of the dispersion (either at $k=0$ or at finite momentum) the amplitudes associated with these localized components become comparable to the amplitudes of the oscillating components, resulting in a measurably enhanced edge-LDOS. The states responsible for this effect are what we call \emph{non-topological edge states}.

Extending the picture presented in this section beyond RWA, one has to deal with a larger number of boundary conditions, while being provided a correspondingly larger number of momenta with non-negative imaginary part. Relevant for the emergence of NTESs are again only those momenta close to a bifurcation point of the dispersion and in this sense the construction laid out above may be viewed as a prototype for the construction of NTESs in higher truncation orders (which of course is unfeasible analytically).

\end{appendix}

\end{document}